\documentclass[twocolumn,showpacs,showkeys,preprintnumbers,superscriptaddress,amsmath,floatfix,amssymb,secnumarabic]{revtex4}

\usepackage{color}
\usepackage[colorlinks=true]{hyperref}
\usepackage{graphicx}
\usepackage{ulem}

\newcommand{\comment}[1]{}

\newcommand{\lr}[1]{ \left( #1 \right) }
\newcommand{\lrs}[1]{ \left[ #1 \right] }
\newcommand{\lrc}[1]{ \left\{ #1 \right\} }
\newcommand{\vev}[1]{ \langle \, #1 \, \rangle }

\newcommand{\tr}{ {\rm tr} \, }

\newcommand{\ket}[1]{ \, | #1 \rangle }
\newcommand{\bra}[1]{ \langle #1 | \, }

\newcommand{\diag}[1]{ {\rm diag} \, \left( #1 \right) }

\renewcommand{\det}[1]{ {\rm det} \left( #1 \right) }

\newcommand{\sign}{ {\rm sign} \,  }

\newcommand{\expa}[1]{ \exp{\left( #1 \right)} }

\newcommand{\dslash}[1]{ #1 \!\!\!/}

\begin{document}
\sloppy

\title{Spontaneous chiral symmetry breaking and the Chiral Magnetic Effect for interacting Dirac fermions with chiral imbalance}

\author{P. V. Buividovich}
\email{pavel.buividovich@physik.uni-regensburg.de}
\affiliation{Institute of Theoretical Physics, University of Regensburg, D-93053 Germany, Regensburg, Universit\"{a}tsstra{\ss}e 31}

\date{August 20, 2014}

\begin{abstract}
 We report on a mean-field study of spontaneous breaking of chiral symmetry for Dirac fermions with contact interactions in the presence of chiral imbalance, which is modelled by nonzero chiral chemical potential. We point out that chiral imbalance lowers the vacuum energy of Dirac fermions, which leads to the increase of the renormalized chiral chemical potential upon chiral symmetry breaking. The critical coupling strength for the transition to the broken phase is slightly lowered as the chiral chemical potential is increased, and the transition itself becomes milder. Furthermore, we study the chiral magnetic conductivity in different phases and find that it grows both in the perturbative weak-coupling regime and in the strongly coupled phase with broken chiral symmetry. In the strong coupling regime the chiral magnetic effect is saturated by vector-like bound states (vector mesons) with mixed transverse polarizations. General pattern of meson mixing in the presence of chiral imbalance is also considered. We discuss the relevance of our study for Weyl semimetals and strongly interacting QCD matter. Finally, we comment on the ambiguity of the regularization of the vacuum energy of Dirac fermions in the presence of chirality imbalance.
\end{abstract}

\maketitle

\section{Introduction}
\label{sec:intro}

 Transport properties of strongly interacting chiral fermions have become a subject of intense research in recent years. One of the fascinating features of chiral fermions is the existence of the so-called anomalous transport phenomena, which stem from quantum anomalies and are thus absent in classical systems \cite{Kharzeev:12:1}. One of the well-known examples of such phenomena is the Chiral Magnetic Effect (CME) - the generation of electric current along the magnetic field in a system with different numbers of left- and right-handed chiral fermions \cite{Kharzeev:08:2}.

 The interest to CME has been to a large extend stimulated by the possibility to observe it in non-central heavy-ion collisions, where the chirality imbalance might be created locally due to topological transitions in the produced quark-gluon plasma and the huge magnetic field with strength comparable to hadronic scale is created due to the relative motion of ions with large electric charge \cite{Kharzeev:08:1}. The CME can also be realized in liquid helium, where its experimental manifestation is the helical instability \cite{VolovikHeliumDroplet}.

 Later on it has been realized that the CME could also be realized in Weyl semimetals \cite{Zyuzin:12:1, Burkov:13:1, Goswami:13:1, Goswami:12:1, Vazifeh:13:1, Zhou:13:1} -- a novel phase of matter in which low-energy excitations are described as Weyl fermions, with left- and right-handed Weyl points being separated either in momentum or in energy \cite{Wan:11:1, Burkov:11:1}. The CME can be observed if the Weyl points of different chiralities have different energies.

 It is a common statement that the transport coefficients which correspond to the CME as well as to other anomalous transport phenomena are universal and do not change when the interactions between fermions are switched on. This statement, however, relies on a number of nontrivial assumptions on the properties of the underlying quantum field theory, such as the existence of a Fermi surface with well-defined quasiparticle excitations around it \cite{Son:12:2, Stephanov:12:1, Son:13:1} or the finiteness of the static screening length \cite{Jensen:12:1, Jensen:12:2, Banerjee:12:1}. It is easy to see that both of these assumptions are violated when spontaneous chiral symmetry breaking occurs. First, the emergence of massless Goldstone bosons leads to the infinite static screening length. Second, the effective mass term generated due to spontaneous symmetry breaking invalidates the Fermi liquid picture at finite chiral chemical potential \cite{Buividovich:13:8}. The generalization of the chiral kinetic equations of \cite{Stephanov:12:1, Son:13:1} to massive fermions was discussed recently in \cite{TorresRincon:13:1, TorresRincon:14:1}, however, this results are still valid only in the realm of the applicability of kinetic theory, that is, for weakly coupled dilute plasmas, for which one do not expect any spontaneous symmetry breaking. Let us also mention that the asymptotic behavior of the chiral magnetic conductivity at high momenta can be related to a certain correlator of two vector and one axial current, which is not renormalized in massless QCD only if the chiral symmetry is not spontaneously broken \cite{Vainshtein:03:1, Knecht:04:1, Buividovich:13:8}.

 While the universal value of the chiral magnetic conductivity can be formally derived from the low-energy chiral Lagrangian \cite{Fukushima:12:1, Kalaydzhyan:14:1, Ray:12:1}, the role of the chiral chemical potential in this derivation is played by the time derivative of the axion field, which makes its interpretation in the Euclidean finite-temperature path integral formalism quite unclear \cite{Vazifeh:13:1}. In particular, it is not clear how to describe the stationary CME current as a response to static magnetic field in such a framework. Since the derivation of \cite{Fukushima:12:1} relies on the QCD chiral Lagrangian, it is also not directly applicable to Weyl semimetals.

 Apart from spontaneous chiral symmetry breaking, anomalous transport coefficients can receive purely radiative corrections if the corresponding currents/charges are coupled to dynamical gauge fields \cite{Jensen:13:1, Miransky:13:1, Gursoy:14:1}. Since in Weyl semimetals the interactions are naturally associated with electric charges, one can expect that even perturbatively the CME current might receive some corrections. Finally, since the chiral chemical potential itself is coupled to a non-conserved axial charge, it might also be subject to some non-trivial renormalization in interacting theories (there are some subtle points in this statement which we address a bit later in this Section).

 In this work we study spontaneous chiral symmetry breaking in a system of interacting Dirac fermions with chiral imbalance, addressing in particular the renormalization of chiral chemical potential and the fate of the Chiral Magnetic Effect in a phase with broken chiral symmetry. Since we mostly keep in mind the application of our results to Dirac quasiparticles in Weyl semimetals, we consider a single flavour of Dirac fermions with instantaneous on-site interactions between electric charges. However, since the consequences of the chiral symmetry breaking are to a large extent independent of the underlying interactions, we believe that our results should be also at least partly applicable to strongly interacting matter and quark-gluon plasma.

 Our main tool in this work will be the mean-field approximation which incorporates possible condensation of all fermionic bilinear operators. While this approximation might eventually break down at sufficiently strong coupling due to large fluctuations of fermionic condensates, we justify our approach by the fact that in most cases mean-field approximation correctly predicts possible patterns of spontaneous symmetry breaking and the types of the phase transitions, while the exact position of the transition points might be incorrect. On the other hand, at weak coupling mean-field approximation simply reproduces an infinite chain of one-loop diagrams.

The chiral imbalance in our study is implemented in terms of a nonzero chiral chemical potential $\mu_A$ coupled to the non-conserved axial charge $Q_A = \bar{\psi} \gamma_0 \gamma_5 \psi$. The term $\mu_A Q_A$ in the single-particle Dirac Hamiltonian shifts the energies of left- and right-handed Weyl nodes to $\pm \mu_A$. We note that while chiral symmetry breaking for Weyl semimetals with spatial momentum separation between the Weyl nodes (which corresponds to the term $b_i \bar{\psi} \gamma_i \gamma_5 \psi$) have been studied in details \cite{Wang:13:1, Wei:12:1, Sekine:13:1, Sukhachov:14:1}, to our knowledge the case of energy separation between Weyl nodes has been previously considered only in the context of effective QCD models \cite{Fukushima:10:1,Fukushima:10:2,Gatto:11:1,Ruggieri:11:1,Nedelin:11:1,Andrianov:13:1,Andrianov:13:2}. In contrast to these works, here we will explore possible fermionic condensates more systematically, in particular taking into account the renormalization of the chiral chemical potential. We will also explicitly take into account the variation of fermionic condensates in external electromagnetic fields which probe the CME. It turns out that this more systematic treatment predicts the enhancement of CME due to interactions, in contrast to the dielectric screening found previously in \cite{Fukushima:10:1}. Another difference of our study from the studies of QCD effective models is that the parameter which controls the breaking of chiral symmetry is the interaction strength rather than the temperature.

 Because of the non-conservation of the axial charge, the chiral chemical potential $\mu_A$ is not a chemical potential in the usual sense. For instance, its value might be renormalized due to interactions. Moreover, it has been shown that in the presence of nonzero $\mu_A$ chiral fermions coupled to electromagnetism become unstable towards the formation of magnetic background with nontrivial Chern-Simons number (also known as magnetic helicity in plasma physics), which effectively reduces the chiral chemical potential \cite{Zamaklar:11:1,Yamamoto:13:1, Sadofyev:13:1}. These facts suggest that the fully self-consistent description of chirally imbalanced matter should be dynamical and should allow for spontaneous breaking of translational invariance. Since such a dynamical description might be quite complicated beyond the kinetic theory/hydrodynamical approximation, here we partly neglect the coupling of fermions to dynamical electromagnetism and assume that there is a spatially homogeneous stationary ground state even in the presence of chiral imbalance.

 In the case of Weyl semimetals, such an approximation might be justified by the fact that in condensed matter systems interactions with magnetic field are suppressed by a factor $v_F^2$ as compared to electrostatic interactions, where $v_F$ is the Fermi velocity (in units of the speed of light). Since the decay of chiral imbalance necessarily involves the generation of magnetic fields, one can expect that the typical decay time will be enhanced by a factor of $1/v_F^2$ as compared to the time scales at which electrostatic interactions are important, and the stationary ground state might be a good approximation at such short time scales. Another possible situation which can be described in terms of the (quasi-)stationary ground state is when chirality is pumped into the system at a constant rate which compensates for its decay rate. Experimental realization of such ``chirality pumping'' in parallel electric and magnetic fields have been recently discussed in \cite{Parameswaran:13:1, Ashby:14:1, Hosur:14:1}.

 The main results of the present work are, first, the quick growth of the renormalized chiral chemical potential in a phase with spontaneously broken chiral symmetry. The corresponding phase transition itself turns into a crossover and is shifted to slightly smaller values of the interaction potential in the presence of chiral imbalance. Second, we find that the CME is significantly enhanced due to interactions, both in the weak- and in the strong-coupling regimes. However, we do not find any discontinuity of the chiral magnetic conductivity across the phase transition. Third, we find that in the strongly-coupled regime the CME current is saturated by the parity-even particle-antiparticle bound state of spin one (or vector meson in QCD terminology). In the presence of chiral imbalance, such states with orthogonal transverse polarizations are mixed with each other, thus giving rise to the parity-odd CME response. Moreover, such vector-like bound states are mixed with pseudo-vector ones. Finally, we comment on very different responses of Dirac fermions to chiral imbalance in the context of condensed matter systems, where the number of states in the Dirac sea is always finite, and in the context of quantum field theories which typically require some regularization of the vacuum energy. While in the former case the vacuum energy is always lowered by chiral imbalance, in the latter case the contribution of regulator fermions leads to the opposite effect.

 The structure of the paper is the following: in Section \ref{sec:hamiltonian} we introduce the model Hamiltonian which we consider and by applying the Hubbard-Stratonovich transformation bring the corresponding partition function into the form suitable for the mean-field calculation. In Section \ref{sec:phase_transition} we study spontaneous breaking of chiral symmetry in the presence of chiral imbalance by numerically minimizing the mean-field free energy. In Section \ref{sec:cme} we study the Chiral Magnetic Effect within the linear response approximation and comment on the mixing between different particle-hole bound states at nonzero chiral chemical potential. In Section \ref{sec:vacuum_energy} we speculate on the role of chiral chemical potential in different regularizations or lattice implementations of Dirac fermions. Finally, in Section \ref{sec:conclusions} we conclude with a general discussion of the obtained results and an outlook for future work.

\section{The model: Dirac Hamiltonian with on-site interactions and chiral chemical potential}
\label{sec:hamiltonian}

 The starting point of our analysis is the many-body Hamiltonian of the following general form:
\begin{eqnarray}
\label{HamiltonianInitial}
 \hat{H} = \hat{H}_0 + \hat{H}_I , \quad
\hat{H}_0 = \sum\limits_{x,y} \hat{\psi}^{\dag}_{x,\alpha} h^{\lr{0}}_{x,\alpha;y,\beta} \hat{\psi}_{y,\beta}  ,
 \nonumber \\
 \hat{H}_I = V \sum\limits_x \hat{q}_x^2   .
\end{eqnarray}
Here $\hat{H}_0$ and $\hat{H}_I$ denote the free and the interaction parts of the Hamiltonian and $\hat{\psi}^{\dag}_{x,\alpha}$ and $\hat{\psi}_{y,\alpha}$ are the fermionic creation and annihilation operators at points $x$ and $y$, which can be either points in continuous space or the sites of some lattice. Correspondingly, the sum $\sum\limits_{x,y}$ denotes either integration over continuous coordinates or summation over lattice sites. Small Greek letters $\alpha$, $\beta$, $\ldots$ label Dirac spinor components. $h^{\lr{0}}_{x,\alpha;y,\beta}$ is the (bare) single-particle Dirac Hamiltonian, which again can be either the continuum Dirac Hamiltonian or some lattice Hamiltonian which at low energies describes Dirac fermions. The bare chiral chemical potential $\mu_A^{\lr{0}}$ is introduced as a term of the form $-\mu_A^{\lr{0}} \lr{\gamma_5}_{\alpha\beta} \delta_{xy}$ in the one-particle Hamiltonian $h_{x,\alpha;y,\beta}$. $V$ is the on-site interaction potential and $\hat{q}_x = \sum\limits_{\alpha} \hat{\psi}^{\dag}_{x,\alpha} \hat{\psi}_{x,\alpha} - 2$ is the charge operator at point $x$. The subtraction of $2$ from the charge operator mimics the background charge of ions in any real crystalline lattice which can support Dirac quasiparticle excitations. This addition, however, does not play any role in our calculation. While at the next stages of our calculations we will use the continuum Dirac Hamiltonian, for the derivation of the mean-field approximation which we present in this Section it is more convenient to assume that $x$ and $y$ in (\ref{HamiltonianInitial}) take discrete values.

 In order to represent the partition function $\mathcal{Z} = \tr\expa{-\beta \hat{H}}$ (where $\beta \equiv T^{-1}$ and $T$ is the temperature) in the form suitable for mean-field calculations, we start with the Suzuki-Trotter decomposition
\begin{eqnarray}
\label{SuzukiTrotter}
 \tr\expa{-\beta \hat{H}}
 = \nonumber \\ =
 \lim\limits_{\Delta\tau \rightarrow 0} \tr\lr{
 e^{-\Delta \tau \hat{H}_0}
 e^{-\Delta \tau \hat{H}_I}
 e^{-\Delta \tau \hat{H}_0}
 e^{-\Delta \tau \hat{H}_I} \ldots
 } ,
\end{eqnarray}
where the Euclidean time interval $\tau \in \lrs{0, \beta}$ is split into infinitely small intervals of size $\Delta \tau$. The representation (\ref{SuzukiTrotter}) is exact up to corrections of order of $O\lr{{\Delta \tau}^2}$. The next step is to apply the Hubbard-Stratonovich transformation to the terms involving the interaction Hamiltonian $\hat{H}_I$. Since charge operators $\hat{q}_x$ in $\hat{H}_I$ commute at different points $x$, we can write
\begin{eqnarray}
\label{InteractionPointSplitting}
 \expa{-\Delta \tau \hat{H}_I} = \prod \limits_x \expa{-\Delta \tau \, V \, \hat{q}_x^2} .
\end{eqnarray}
There are now several possibilities to perform the Hubbard-Stratonovich transformation on each of the factors on the right-hand side of (\ref{InteractionPointSplitting}), corresponding to different grouping of four fermionic operators in (\ref{InteractionPointSplitting}) into two fermionic bilinears. If we were able to perform the integration over the Hubbard field exactly, all these representations would be of course equivalent. However, the mean-field approach which we use in this work is the saddle-point approximation for the integral over the Hubbard field, and the validity of this approximation might strongly depend on the choice of the integration variables. Here we perform the Hubbard-Stratonovich transformation in a way which allows to treat the chiral condensate and also all other fermionic bilinear condensates as extrema of the corresponding effective action. To this end we first rewrite
\begin{eqnarray}
\label{q2transform1}
 \hat{q}_x^2 = \lr{\hat{\psi}^{\dag}_{x,\alpha} \hat{\psi}_{x,\alpha} - 2} \lr{\hat{\psi}^{\dag}_{x,\beta} \hat{\psi}_{x,\beta} - 2}
 = \nonumber \\ =
 -\hat{\psi}^{\dag}_{x,\alpha} \hat{\psi}_{x,\beta} \hat{\psi}^{\dag}_{x,\beta} \hat{\psi}_{x,\alpha} + \hat{\psi}^{\dag}_{x,\alpha} \hat{\psi}_{x,\alpha} ,
\end{eqnarray}
where from now on we shorten the notation by assuming summation over repeated spinor indices. Inserting this representation of $\hat{q}_x^2$ into (\ref{InteractionPointSplitting}), we can separate the two summands in the second line of (\ref{q2transform1}) into two exponents, making an irrelevant error of order $O\lr{{\Delta \tau}^2}$. Then we transform the exponent containing the four fermionic operators into the exponent of a fermionic bilinear operator by representing it in terms of a Gaussian integral over the Hubbard-Stratonovich field $\Phi_{x,\alpha\beta}$, which is a Hermitian matrix in spinor space:
\begin{widetext}
\begin{eqnarray}
\label{q2transform2}
 \expa{V {\Delta \tau} \hat{\psi}^{\dag}_{x,\alpha} \hat{\psi}_{x,\beta} \hat{\psi}^{\dag}_{x,\beta} \hat{\psi}_{x,\alpha}}
 = 
 \int d\Phi_{x,\alpha\beta} \expa{- \frac{\Delta \tau}{4 V} \Phi_{x,\alpha\beta}\Phi_{x,\beta\alpha}  - {\Delta \tau} \Phi_{x,\alpha \beta} \hat{\psi}^{\dag}_{x,\alpha} \hat{\psi}_{x,\beta} }  .
\end{eqnarray}
Finally, we can again combine all the factors in the Suzuki-Trotter decomposition (\ref{SuzukiTrotter}) into a single time-ordered exponent, neglecting the error of order of $O\lr{{\Delta \tau}^2}$, which yields
\begin{eqnarray}
\label{HSPartitionFunction}
 \mathcal{Z} = \int \mathcal{D}\Phi_{x,\alpha \beta}\lr{\tau}
 \expa{ - \frac{1}{4 V} \int\limits_{0}^{\beta} d\tau \sum\limits_x \Phi_{x,\alpha\beta}\lr{\tau} \Phi_{x,\beta\alpha}\lr{\tau}}
 \times \nonumber \\ \times
 \tr \mathcal{T} \expa{-\int\limits_{0}^{\beta} d\tau \lr{\hat{H}_0 + \sum\limits_{x} \lr{\Phi_{x,\alpha\beta}\lr{\tau} + V \delta_{\alpha\beta}} \hat{\psi}^{\dag}_{x,\alpha} \hat{\psi}_{x,\beta} }}    .
\end{eqnarray}
\end{widetext}
We thus have formulated the problem in terms of the fermionic partition function which corresponds to the effective time-dependent single-particle Hamiltonian of the following form:
\begin{eqnarray}
\label{EffectiveSPHamiltonian}
 h_{x,\alpha;y,\beta} = h^{\lr{0}}_{x,\alpha;y,\beta} + V \delta_{xy} \delta_{\alpha\beta} + \Phi_{x,\alpha\beta}\lr{\tau} \delta_{xy}  ,
\end{eqnarray}
with background field $\Phi_{x,\alpha\beta}\lr{\tau}$ which depends on Euclidean time $\tau$.

 In the mean-field approximation, we replace the integral over $\Phi_{x,\alpha\beta}\lr{\tau}$ in (\ref{HSPartitionFunction}) by the value of the integrand at its minimum. If the minimum corresponds to some nonzero value of $\Phi_{x,\alpha\beta}\lr{\tau}$, one says that the fermionic condensate $\vev{\hat{\psi}^{\dag}_{x,\alpha} \hat{\psi}_{x,\beta}} \sim \Phi_{x,\alpha\beta}$ is formed. Additionally, one can take into account the corrections by integrating over Gaussian fluctuations around the saddle point, which are interpreted as propagating bound states of two fermions.

 In order to find the minimum, we will assume that the invariance under shifts of Euclidean time is not broken, and thus the value of the Hubbard-Stratonovich field $\Phi_{x,\alpha\beta}\lr{\tau} \equiv \Phi_{x,\alpha\beta}$ at the minimum does not depend on $\tau$. Then the trace of the time-ordered exponent in the integrand in (\ref{HSPartitionFunction}) can be rewritten as simply the partition function of a free fermion gas with single-particle Hamiltonian (\ref{EffectiveSPHamiltonian}) with $\tau$-independent $\Phi_{x,\alpha\beta}$:
\begin{eqnarray}
\label{FermionicTrace}
\tr \expa{-\beta \sum\limits_{x,y,\alpha,\beta} \hat{\psi}^{\dag}_{x,\alpha} h_{x,\alpha;y,\beta} \hat{\psi}_{y,\beta}}
 = \nonumber \\ =
 \expa{\sum\limits_i \log\lr{1 + e^{-\epsilon_i/T}} }   ,
\end{eqnarray}
where the sum goes over all energy levels $\epsilon_i$ of this single-particle Hamiltonian. In this work we will be interested in the limit of zero temperature. In this case all the summands in the second line of (\ref{FermionicTrace}) with $\epsilon_i > 0$ are zero, and all the terms with $\epsilon_i < 0$ are equal to $\epsilon_i/T$. We thus see that the fermionic contribution to the free energy $\mathcal{F} = -T \log\mathcal{Z}$ is simply equal to the (negative) energy of the Dirac sea, which is just the sum of all the energy levels below zero. Combining this contribution with the quadratic action of the Hubbard-Stratonovich field, we conclude that in the static mean-field approximation we have to minimize the following functional with respect to $\Phi_{x,\alpha\beta}$:
\begin{eqnarray}
\label{MFFunctional}
 \mathcal{F} = \sum\limits_{\epsilon_i<0} \epsilon_i
  +
 \sum\limits_x \frac{\Phi_{x,\alpha\beta}\Phi_{x,\beta\alpha}}{4 V} .
\end{eqnarray}

 Before proceeding with the actual minimization of the above functional, let us consider the effective chemical potential term $V \delta_{\alpha\beta} \delta_{xy}$ in the effective single-particle Hamiltonian (\ref{EffectiveSPHamiltonian}). Its appearance might seem strange at the first sight, since it might induce nonzero electric charge in an initially electrically neutral system. However, the Hubbard-Stratonovich field $\Phi_{x,\alpha\beta}$ also contains the constant component proportional to $\delta_{\alpha\beta}$ which can mimic the chemical potential. Let us explicitly separate this term by writing $\Phi_{x,\alpha\beta} = \mu \delta_{\alpha\beta} + \tilde{\Phi}_{x,\alpha\beta}$, with $\sum\limits_x \tilde{\Phi}_{x,\alpha\alpha} = 0$. The functional (\ref{MFFunctional}) can be then written as
\begin{eqnarray}
\label{MFFunctionalMu}
 \mathcal{F} = \sum\limits_{\tilde{\epsilon}_i<-\mu-V} \lr{\tilde{\epsilon}_i + \mu + V}
 + \nonumber \\ +
 \sum\limits_x \frac{\tilde{\Phi}_{x,\alpha\beta} \tilde{\Phi}_{x,\beta\alpha}}{4 V} + \frac{\mu^2 L^3}{V} ,
\end{eqnarray}
where $L$ is the spatial size of the system and $\tilde{\epsilon}_i$ are the eigenvalues of the single-particle Hamiltonian (\ref{EffectiveSPHamiltonian}) with $\Phi_{x,\alpha\beta}$ replaced by $\tilde{\Phi}_{x,\alpha\beta}$ and without the $V \delta_{\alpha\beta} \delta_{xy}$ term. In order to find the mean-field value of $\mu$, we have to solve the equation $\frac{\partial}{\partial \mu} \mathcal{F} = 0$, which can be written as
\begin{eqnarray}
\label{MuMinimization}
 \sum\limits_i \theta\lr{-\tilde{\epsilon}_i - \mu - V} + 2 L^3 \mu/V = 0  ,
\end{eqnarray}
where $\theta$ is the Heaviside step function, which simply counts the number of energy levels $\epsilon_i = \tilde{\epsilon}_i + \mu + V$ below zero. One can show that for Dirac Hamiltonians with chiral chemical potential either in the continuum or on the lattice the energy levels $\tilde{\epsilon}_i$ are symmetric around zero, so that every positive energy level $\tilde{\epsilon}_i > 0$ is accompanied by the negative energy level $-\tilde{\epsilon}_i$. The total number of energy levels is $N = 4 L^3$ and is equal to the total number of degrees of freedom living on the lattice ($4$-component Dirac spinors on every lattice site). Therefore there are $2 L^3$ energy levels $\tilde{\epsilon}_i$ below zero. Taking this observation into account, it is easy to see that $\mu = -V$ is the stable solution of the equation (\ref{MuMinimization}). We conclude therefore, that the effective mean-field chemical potential exactly compensates the term $V \delta_{\alpha\beta} \delta_{xy}$ in (\ref{EffectiveSPHamiltonian}). Therefore we can simply discard this term and assume that the Hubbard-Stratonovich field $\Phi_{x,\alpha\beta}$ is traceless on average: $\sum\limits_x \Phi_{\alpha\alpha} = 0$.

\section{Mean-field study of spontaneous chiral symmetry breaking in the presence of chiral imbalance}
\label{sec:phase_transition}

 In this Section we study the phase diagram of our model by explicitly minimizing the free energy (\ref{MFFunctional}). In what follows, we will use the continuum regularized Dirac Hamiltonian for the effective single-particle Hamiltonian $h_{x,\alpha; y,\beta}$ in (\ref{EffectiveSPHamiltonian}):
\begin{eqnarray}
\label{ContinuumDiracHamiltonian}
 h_{x;y} = \lr{-i v_F \alpha_i \nabla^x_i + \mu^{\lr{0}}_A \gamma_5 + \Phi_{x}} \tilde{F}\lr{x - y, \Lambda} ,
\end{eqnarray}
where $v_F$ is the Fermi velocity, $x$ and $y$ are now the continuum coordinates, $\nabla^x_i = \partial_{x,i} + i A_{x,i}$ is the covariant derivative over $x$ which includes external gauge field $A_{x,i}$ and $\tilde{F}\lr{x - y, \Lambda}$ is some ultraviolet regularization of the delta-function $\delta\lr{x - y}$, which very quickly decays if $x$ and $y$ are separated by the distance larger than the inverse UV cutoff scale $\Lambda^{-1}$. $\alpha_i = \gamma_0 \gamma_i = \diag{\sigma_i, -\sigma_i}$ are the Dirac $\alpha$-matrices, $\gamma_5 = \diag{I, -I}$ is the generator of chiral rotations, $\gamma_{\mu}$ are the Dirac gamma-matrices, $\sigma_i$ are the Pauli matrices and $I$ is the $2 \times 2$ identity matrix. We also suppress the spinor indices of the Dirac matrices and $\Phi_{x,\alpha\beta}$ for the sake of brevity.

 The main reason to use the continuum Dirac operator in this exploratory study is to preserve exact chiral symmetry of the action. First of all, this makes our results potentially applicable to a wider range of physical systems, including also QCD. Second, the mean-field calculations in the continuum approximation are much simpler and also more instructive. While for more realistic lattice Dirac Hamiltonians the chiral symmetry is always broken at high energies due to lattice artifacts \footnote{Except for the case of the overlap Hamiltonian proposed in \cite{Creutz:01:1}, implementation of which would be, however, a fantastic fine-tuning in any real crystal}, we expect that the continuum approximation which we make here nevertheless captures some features of the low-energy behavior of lattice models. Preliminary studies with Wilson-Dirac Hamiltonian, which will be published elsewhere, qualitatively confirm the predictions obtained in the continuum approximation.

 In order to calculate the mean-field diagram for the effective single-particle Hamiltonian (\ref{ContinuumDiracHamiltonian}), we assume that external fields are absent and translational and rotational symmetries are not broken, so that $\Phi_{x,\alpha\beta} \equiv \Phi_{\alpha\beta}$ is constant in space. Moreover, Lorentz symmetry allows only the following structure of fermionic condensates $\vev{\hat{\psi}^{\dag}_{x,\alpha} \hat{\psi}_{x,\beta}} \sim \Phi_{\alpha\beta}$:
\begin{eqnarray}
\label{CondensateStructure}
 \Phi
 =
 m \cos{\theta} \gamma_0 + i m \sin{\theta} \gamma_0 \gamma_5 + \gamma_5 \lr{\mu_A - \mu_A^{\lr{0}}}  ,
\end{eqnarray}
where $m$ is the effective mass induced due spontaneous chiral symmetry breaking, $\theta$ is the complex phase of the mass and $\mu_A$ is the renormalized chiral chemical potential. Further we will see that the mean-field free energy does not depend on $\theta$, which is thus the progenitor of a Goldstone mode. It is easy to calculate the energy levels $\epsilon_{s,\sigma}\lr{\vec{k}}$ of the effective single-particle Hamiltonian (\ref{EffectiveSPHamiltonian}) with such Hubbard-Stratonovich field $\Phi$:
\begin{eqnarray}
\label{MuAEnergyLevels}
 \epsilon_{s,\sigma}\lr{\vec{k}} = \varepsilon_{s,\sigma}\lr{\vec{k}} \, F\lr{\vec{k}, \Lambda},
 \nonumber \\
 \varepsilon_{s,\sigma}\lr{\vec{k}} = s \sqrt{\lr{v_F |\vec{k}| - \sigma \mu_A}^2 + m^2}
\end{eqnarray}
where $s, \, \sigma = \pm 1$, $F\lr{\vec{k}, \Lambda} = \int d^3 x e^{i \vec{k} \cdot \vec{x}} F\lr{\vec{x}, \Lambda}$ is the factor which imposes UV cutoff and $\varepsilon_{s,\sigma}\lr{\vec{k}}$ denotes the energy levels of unregularized continuum Dirac Hamiltonian. In what follows, we assume that the function $F\lr{\vec{k}, \Lambda}$ is equal to one for $|\vec{k}| < \Lambda$ and for $|\vec{k}| > \Lambda$ it approaches zero very quickly, so that for sufficiently smooth integrands $A\lr{\vec{k}}$ the integrals of the form $\int d^3 k F\lr{\vec{k}, \Lambda} A\lr{\vec{k}}$ can be written as $\int\limits_{|\vec{k}|<\Lambda} d^3 k A\lr{\vec{k}}$. Such UV regularization makes the energy of the Dirac sea (the first summand in the mean-field free energy (\ref{MFFunctional})) finite and mimics the finiteness of the Dirac sea in more realistic lattice models, for which $\Lambda$ can be associated with inverse lattice spacing.

 In continuum space the action of the Hubbard field, which is the second part of the mean-field functional (\ref{MFFunctional}), can be in general written as
\begin{eqnarray}
\label{HSActionContinuum}
 \frac{1}{4 V} \sum\limits_x \Phi_{x \, \alpha\beta} \Phi_{x \, \beta\alpha} =
 \frac{c \Lambda^3}{4 V} \int{d^3 x} \, \Phi_{x \, \alpha\beta} \Phi_{x \, \beta\alpha}  ,
\end{eqnarray}
where $c$ is some constant which depends on the details of UV regularization. This formula can be obtained, e.g., by taking the naive continuum limit in some lattice regularization. Clearly, the only effect of the constant $c$ is some renormalization of the interaction potential $V$.

 Inserting now the expressions (\ref{MuAEnergyLevels}), (\ref{CondensateStructure}) and (\ref{HSActionContinuum}) into (\ref{MFFunctional}), we obtain the following expression for the mean-field free energy which should be minimized with respect to the effective mass $m$ and the renormalized chiral chemical potential $\mu_A$:
\begin{eqnarray}
\label{MFFunctionalPractical}
 \frac{\mathcal{F}}{L^3}
 =
 - \int\limits_{|\vec{k}|<\Lambda} \frac{d^3 k}{\lr{2 \pi}^3}
 \sum\limits_{\sigma=\pm 1} \sqrt{\lr{v_F |\vec{k}| - \sigma \mu_A}^2 + m^2}
 + \nonumber \\ +
 \frac{c \Lambda^3}{V} \lr{m^2 + \lr{\mu_A - \mu_A^{\lr{0}}}^2}
\end{eqnarray}

 In order to study spontaneous chiral symmetry breaking, we perform numerical minimization of this functional. To this end we set $c = 1$ and $\Lambda = 1$ in what follows, thus expressing all other quantities in units of UV cutoff. Instead of fixing the Fermi velocity $v_F$ to some particular value, it is convenient to rescale the mean-field free energy as $\bar{\mathcal{F}} = \mathcal{F}/v_F$ and express it in terms of the rescaled fermionic condensates and interaction potential $\bar{m} = m/v_F$, $\bar{\mu}_A = \mu_A/v_F$, $\bar{\mu}_A^{\lr{0}} = \mu_A^{\lr{0}}/v_F$ and $\bar{V} = V/v_F$. It is easy to check that in terms of these new variables, the mean-field free energy (\ref{MFFunctionalPractical}) does not depend on $v_F$:
\begin{eqnarray}
\label{MFFunctionalRescaled}
 \frac{\bar{\mathcal{F}}}{L^3}
 =
 - \int\limits_{|\vec{k}|<\Lambda} \frac{d^3 k}{\lr{2 \pi}^3}
 \sum\limits_{\sigma=\pm 1} \sqrt{\lr{|\vec{k}| - \sigma \bar{\mu}_A}^2 + \bar{m}^2}
 + \nonumber \\ +
 \frac{c \Lambda^3}{\bar{V}} \lr{\bar{m}^2 + \lr{\bar{\mu}_A - \bar{\mu}_A^{\lr{0}}}^2}  .
\end{eqnarray}
We thus see that after minimizing the above expression the dependence of the final results on the Fermi velocity can be obtained by a simple rescaling.

\begin{figure*}[htpb]
  \centering
  \includegraphics[width=6cm,angle=-90]{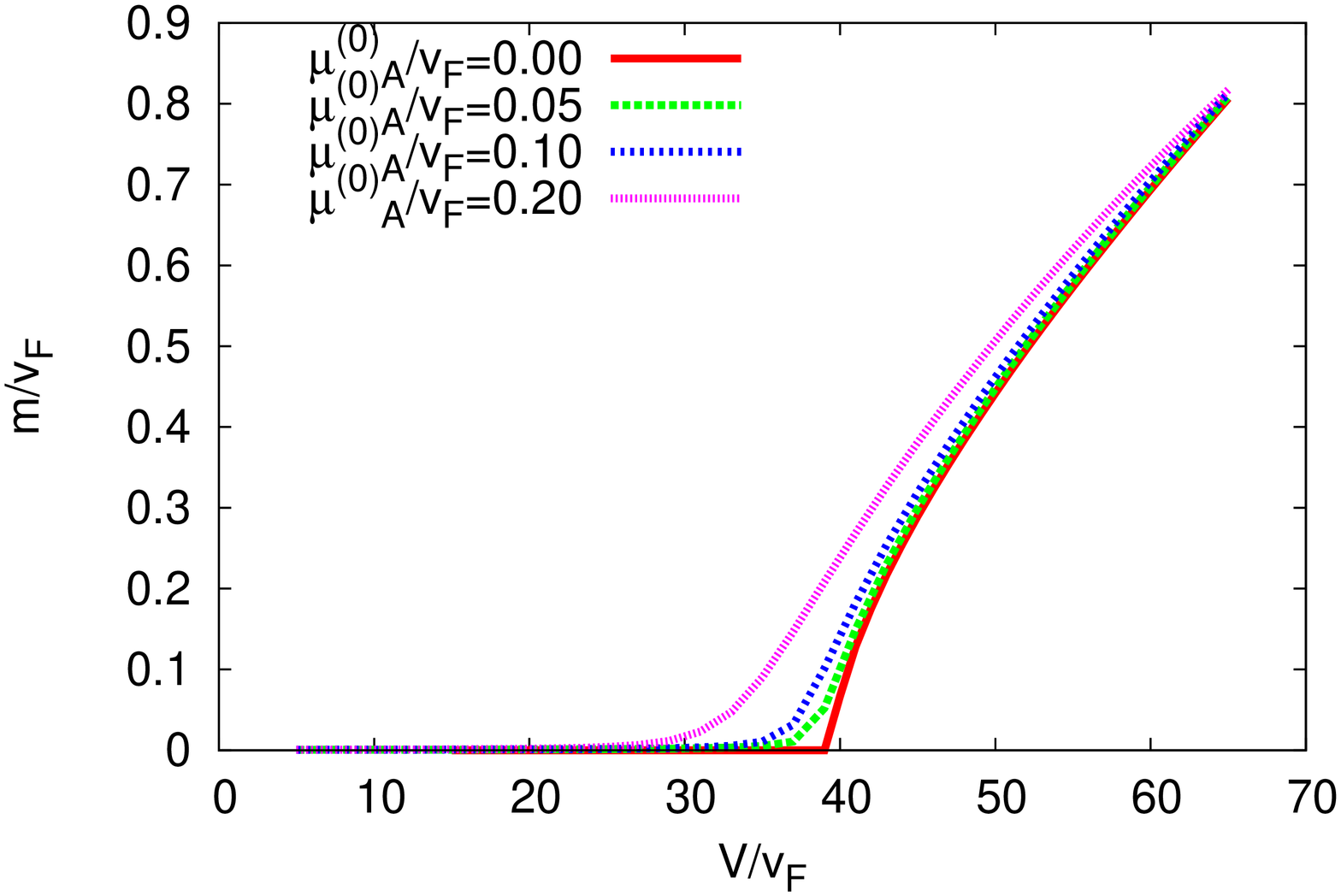}
  \includegraphics[width=6cm,angle=-90]{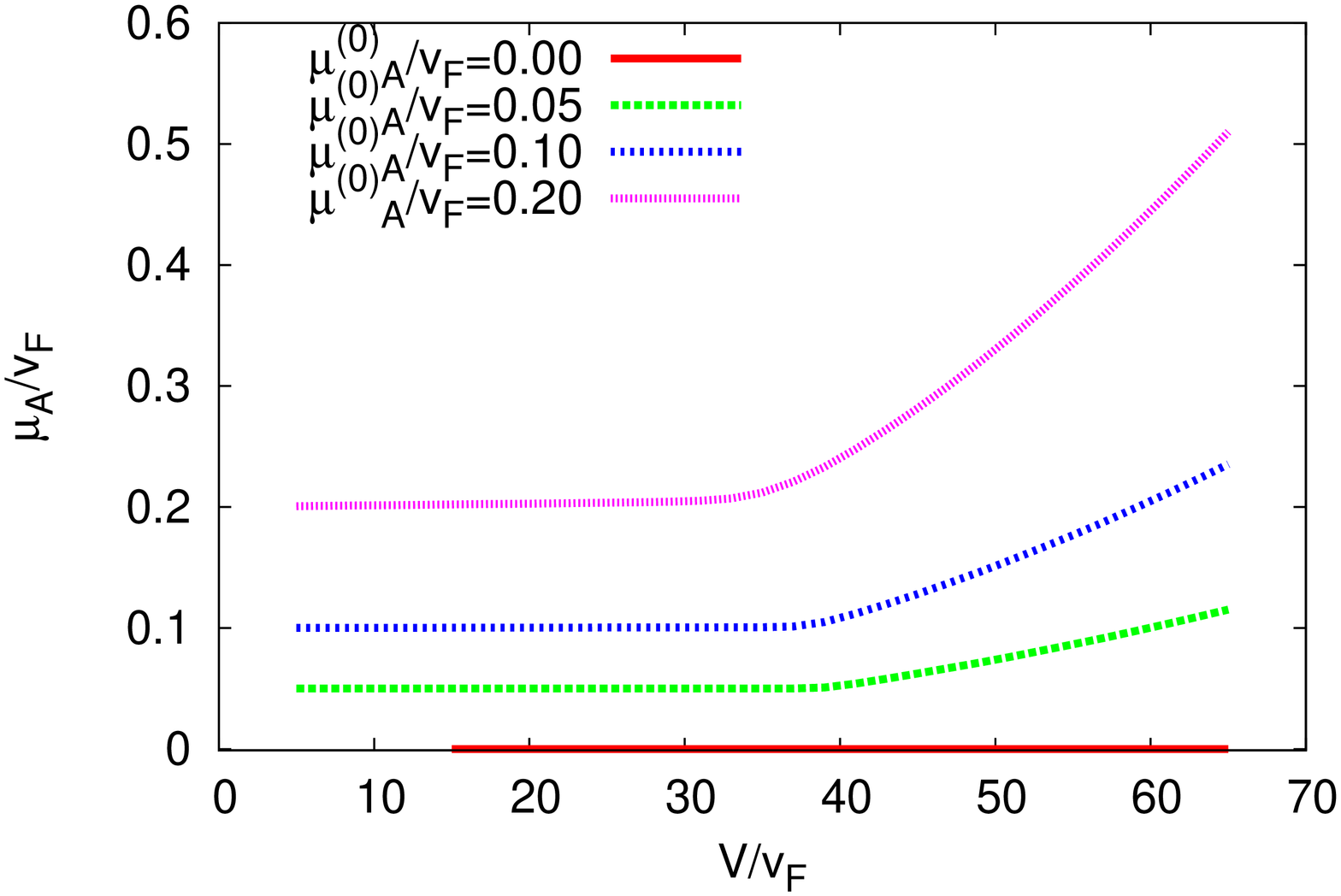}
  \caption{Mean-field values of the effective mass (on the left) and the renormalized chiral chemical potential (on the right) as functions of interaction potential $V$ at different values of the bare chiral chemical potential $\mu_A^{\lr{0}}$. All numbers are given in units of the UV cutoff scale $\Lambda$.}
  \label{fig:mf_condensates}
\end{figure*}

 Saddle point values of the effective mass $m$ and the renormalized chiral chemical potential $\mu_A$ are plotted on Fig.~\ref{fig:mf_condensates} as functions of the interaction potential $V$ for different values of the bare chiral chemical potential $\mu_A^{\lr{0}}$. If the chiral chemical potential is absent in the bare Hamiltonian, we observe the standard picture of second-order quantum phase transition associated with spontaneous chiral symmetry breaking. Namely, the effective mass is identically zero for $V/v_F < V_c/v_F = 39.5$ and rapidly grows at $V > V_c$. The chiral chemical potential is identically zero for all values of $V$ in this case.

 However, at nonzero bare chiral chemical potential $\mu_A^{\lr{0}}$ the situation becomes more interesting. First, the second-order transition changes to some sort of crossover or an infinite-order phase transition already at very small values of $\mu_A^{\lr{0}}$, so that the effective mass $m\lr{V}$ still quickly rises for $V$ larger than some pseudo-critical value but does not show any discontinuity of derivatives. Rather, $m\lr{V}$ very slowly approaches zero as $V$ goes to zero. The renormalized chiral chemical potential $\mu_A$ is very close to the bare value for $V \lesssim V_c\lr{\mu_A^{\lr{0}}}$ but starts quickly growing at $V > V_c\lr{\mu_A^{\lr{0}}}$. The derivatives of $\mu_A\lr{V}$ also do not exhibit any discontinuities. As $\mu_A^{\lr{0}}$ increases, the transition becomes more and more soft.

\begin{figure}[htpb]
  \centering
  \includegraphics[width=6.5cm,angle=-90]{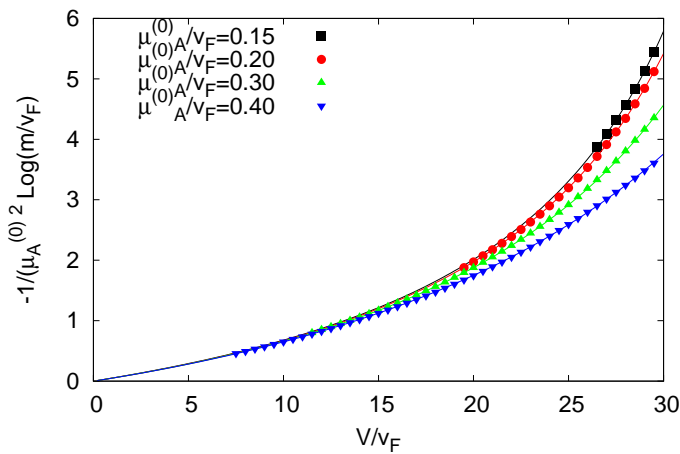}\\
  \caption{Inverse logarithm of the effective mass $m$ at small values of the inter-electron interaction potential $V$. Solid lines are fits to the essential singularity scaling (\ref{BCSGap}).}
  \label{fig:mass_vs_V_logscale}
\end{figure}

 The very slow monotonic growth of the effective mass $m\lr{V}$ at small $V$ is of course expectable, since finite particle-like and hole-like Fermi surfaces at $|k| = |\mu_A|$ trigger Cooper-type instability towards the formation of particle-hole bound states even at arbitrarily small inter-electron interaction potential $V$. The effective mass in this case should have an essential singularity in $V$ of the form
\begin{eqnarray}
\label{BCSGap}
 m = A V^{\alpha}\expa{-\frac{B}{\mu_A^2 V}} ,
\end{eqnarray}
where the factor $\mu_A^2$ is proportional to the density of states at the Fermi level \footnote{For the linear dispersion relation $\epsilon\lr{\vec{k}} = \pm |\vec{k}|$, $\rho\lr{k_F} \sim k_F^2 \sim \mu_A^2$, where $k_F$ is the Fermi momentum.} (see e.g. Chapter 6 of \cite{AltlandSimonsCondMatQFT}) and $A$, $\alpha$ and $B$ are some constants. In order to check this essential singularity scaling of the effective mass, on Fig.~\ref{fig:mass_vs_V_logscale} we plot the inverse logarithm of the effective mass $-1/\log\lr{m}$ rescaled by $\mu_A^2$ as a function of $V$. Due to finite precision of our numerical minimization procedure the effective mass $m$ contains large systematic numerical errors at sufficiently small $V$, and for smaller $\mu_A^{\lr{0}}$ the numerical errors typically set in at larger values of $V$. For this reason we only plot $m$ in the range of $V$ where numerical errors are negligible. The results of numerical minimization are shown on Fig.~\ref{fig:mass_vs_V_logscale} with points. One can see that indeed $-1/\log\lr{m}$ tends to zero in the limit of zero $V$. In order to quantify this tendency, we fit the $V$ dependence of $-1/\log\lr{m}$ with the function $\frac{\mu_A^2 V}{c - \mu_A^2 V \log A - \alpha \mu_A^2 V \log V}$, as suggested by the scaling law (\ref{BCSGap}). These fits are shown on Fig.~\ref{fig:mass_vs_V_logscale} with solid lines. We find that the results of numerical minimization are indeed well fitted by such functions, with the values of the parameters $c$, $A$ and $\alpha$ coinciding within the fitting uncertainty for all values of $\mu_A^{\lr{0}}$.

 We therefore conclude that due to Cooper instability the second-order phase transition at $\mu_A^{\lr{0}} = 0$ turns into a crossover at $\mu_A^{\lr{0}} > 0$, and the true chiral phase transition at which the effective mass $m$ becomes different from zero is shifted to $V = 0$. Since in realistic lattice models the chiral symmetry is anyway broken by the nonlinearity of the lattice dispersion relation away from the Dirac points and the pattern of spontaneous symmetry breaking might become quite different from the one in the continuum theory, we do not study here this crossover transition in more detail. Let us only mention that recently we have performed a similar mean-field study of the phase diagram for Wilson-Dirac Hamiltonian with chiral imbalance, where the chiral phase transition is replaced by the transition to the Aoki phase with condensed pions \cite{Buividovich:14:2}. It turned out that in this case the order of the phase transition does not change in the presence of chiral imbalance, and there are no indications of Cooper-type instabilities. Otherwise, all the qualitative results of \cite{Buividovich:14:2} are similar to those obtained in the present paper.

 For crossover transitions the usual way to define the observable-dependent critical interaction potential is to associate it with the inflection point of the approximate order parameter. From Fig.~\ref{fig:mf_condensates} one can see that the critical coupling $V_c\lr{\mu_A^{\lr{0}}}$ defined as the inflection point of the effective mass becomes smaller as the bare chiral chemical potential grows. This finding is in agreement with the results obtained from QCD-inspired effective models, for which the deconfinement temperature becomes lower at nonzero $\mu_A$ \cite{Fukushima:10:1, Fukushima:10:2, Gatto:11:1, Ruggieri:11:1, Nedelin:11:1}. Let us also note that the inflection point of the derivative $d\mu_A/dV$ can be also used to define the ``critical'' interaction potential, which turns out to be slightly different from the critical value obtained from the inflection point of the effective mass, but also decreases for larger $\mu_A^{\lr{0}}$. This difference in the critical values of $V$ obtained from different observables also indicates that in the presence of chiral chemical potential spontaneous breaking of chiral symmetry is a crossover, at least in the mean-field approximation.

 We see that in contrast to the conventional chemical potential, which is not renormalized by virtue of conservation of electric charge (see the discussion at the end of Section~\ref{sec:hamiltonian}), chiral chemical potential is effectively enhanced by interactions. A simple intuitive explanation of this observation is that the chiral imbalance, much like the Dirac mass term, lowers the vacuum energy and is thus energetically favourable. We discuss this property in more details for both continuum and lattice Dirac fermions in Section \ref{sec:vacuum_energy}.

\section{Chiral Magnetic Effect in the linear response approximation}
\label{sec:cme}

 In this Section we study how the chiral symmetry breaking discussed in the previous Section \ref{sec:phase_transition} affects the CME. Here we consider CME as a stationary process, that is, a response of static, permanently flowing electric current to a static magnetic field. On the one hand, this approximation might be too idealistic to be realized in real physical systems, for which chirality pumping and the application of external magnetic field are always dynamical processes \cite{Burkov:13:1, Parameswaran:13:1, Ashby:14:1, Hosur:14:1}. On the other hand, it is probably the possibility to have charge transport in the ground state of the system which makes the idea of CME so attractive. Thus we assume that such a stationary regime can be at least approached under the same assumptions under which the system of Dirac fermions with chiral imbalance can be considered as a stationary state (see the discussion in the introductory Section \ref{sec:intro}).

 In the linear response approximation, one can characterize such a static CME response by the chiral magnetic conductivity $\sigma_{CME}\lr{\vec{k}}$, where $\vec{k}$ is the wave vector which describes the spatial modulation of sufficiently weak magnetic field which causes the CME. In order to match the hydrodynamical description of the chirally imbalanced plasma one has to take the limit $\vec{k} \rightarrow 0$. We assume that $\vec{k}$ is parallel to the third coordinate axis: $\vec{k} = k_3 \vec{e}_3$. Then $\sigma_{CME}\lr{k_3}$ can be found from the following Kubo formula \cite{Gynther:10:1, Landsteiner:11:2, Landsteiner:12:1}:
\begin{eqnarray}
\label{CME_Kubo}
 \sigma_{CME}\lr{k_3}
 =
 -\frac{i}{k_3} \, \frac{1}{L^3} \sum\limits_{x, y} e^{i k_3 \lr{x_3 - y_3}} \bra{0} \hat{j}_{x,1} \hat{j}_{y,2} \ket{0}
 = \nonumber \\ =
 \left.
 -\frac{i}{k_3} \, \frac{1}{L^3} \sum\limits_{x, y} e^{i k_3 \lr{x_3 - y_3}}
 \frac{\delta^2 \mathcal{F} }{\delta A_{x,1} \delta A_{y,2}} \,
  \right|_{A_{x,i} = 0}
\end{eqnarray}
where $\hat{j}_{x,i} = \hat{\psi}^{\dag}_x \alpha_i \hat{\psi}_x$ is the current density operator, $A_{x,i}$ is the external Abelian gauge field and $\mathcal{F}\lrs{A_{x,i}} = -T \log{\mathcal{Z}\lrs{A_{x,i}}}$ is the free energy of the system which depends on the external gauge field $A_{x,i}$.

 In the mean-field approximation the free energy $\mathcal{F}\lrs{A_{x,i}}$ is approximated by its value at some saddle-point value of the Hubbard field $\Phi_{x,\alpha \beta}$, which we denote as $\Phi_{x,\alpha \beta}^{\star}\lrs{A_{x, i}}$:
\begin{eqnarray}
\label{mf_free_energy_lrt}
 \mathcal{F}\lrs{A_{x,i}} = \mathcal{F}\lrs{\Phi_{x,\alpha \beta}^{\star}\lrs{A_{x,i}}, A_{x,i}}  ,
\end{eqnarray}
where the saddle-point value $\Phi_{x,\alpha \beta}^{\star}\lrs{A_{x,i}}$ is the solution of the equation
\begin{eqnarray}
\label{mf_equation_lrt}
 \frac{\partial}{\partial \Phi_{x,\alpha \beta}} \mathcal{F}\lrs{\Phi_{x,\alpha \beta}, A_{x,i}} = 0  \end{eqnarray}
and $\mathcal{F}\lrs{\Phi_{x,\alpha \beta}, A_{x,i}}$ is now the mean-field free energy (\ref{MFFunctional}) in the presence of external magnetic field $\vec{B} = \vec{\nabla} \times \vec{A}$ described by the gauge vector field $A_{x,i}$. Since we consider the static CME response, all quantities in these equations are completely time-independent. Note also that since the mean-field free energy (\ref{MFFunctional}) in general depends on the external electromagnetic field, the saddle-point value of the Hubbard field $\Phi_{x,\alpha \beta}^{\star}\lrs{A_{x,i}}$ also becomes some functional of the gauge field $A_{x,i}$. This is simply the reflection of the change of fermionic condensates in external gauge field $A_{x,i}$. The variations of the extremum value of the mean-field free energy $\mathcal{F}\lrs{A_{x,i}}$ with $A_{x,i}$ should then take into account both the variation of the functional $\mathcal{F}\lrs{\Phi_{x,\alpha \beta}, A_{x,i}}$ itself and the variation of the saddle-point value of the Hubbard-Stratonovich field $\Phi_{x,\alpha \beta}^{\star}\lrs{A_{x,i}}$, which is in general spatially inhomogeneous. Since the spinor structure of the fermionic condensates in the presence of external gauge field might be quite complicated, for further notational convenience we decompose the spinor part of $\Phi_{x,\alpha\beta}$ in the full basis $\Gamma_A$, $A = 1 \ldots 16$ of hermitian spinor operators:
\begin{eqnarray}
\label{PhiDecomposition}
 \Phi_{x,\alpha\beta} = \sum\limits_{A=1}^{16} \Phi_{x,A} \Gamma_{A,\alpha\beta},
 \nonumber \\
 \Gamma_A = \left\{I \otimes I, I \otimes \sigma_i,
 \tau_1 \otimes I, \tau_1 \otimes \sigma_i,
 \right. \nonumber \\ \left.
 \tau_2 \otimes I, \tau_2 \otimes \sigma_i,
 \tau_3 \otimes I, \tau_3 \otimes \sigma_i \right\}, \quad i = 1,2,3 ,
\end{eqnarray}
where the first factors in the direct products are the matrices with chiral indices $L\lr{eft}$ and $R\lr{ight}$, the second factors are the matrices with spin indices $\uparrow$, $\downarrow$ and $\tau_i$, $i = 1,2,3$ are the Pauli matrices with chiral indices $L$, $R$. We will also use the notation $\Gamma_A \equiv \tau_A \otimes \sigma_A$, where depending on the value of $A$ $\tau_A$ and $\sigma_A$ can be the corresponding Pauli matrices or the identity matrices. The matrices $\Gamma_A$ are normalized as $\tr\lr{\Gamma_A \Gamma_B} = 4 \delta_{AB}$, so that the action of the Hubbard-Stratonovich field (second summand in the mean-field free energy (\ref{MFFunctional})) reads $\frac{1}{4 V} \sum\limits_{x} \Phi_{x,\alpha\beta} \Phi_{x,\beta\alpha} = \frac{1}{V} \sum\limits_{x} \Phi_{x,A}^2$.

 After some manipulations with derivatives of implicit functions, which are summarized in Appendix \ref{apdx:lrt_mf}, we arrive at the following general expression for the second-order variation of the mean-field free energy with respect to external gauge field $A_{x,i}$:
\begin{eqnarray}
\label{mf_response}
 \frac{\delta^2 \mathcal{F}}{\delta A_{x,i} \, \delta A_{y,j}}
 =
 \frac{\partial^2 \mathcal{F}}{\partial A_{x,i} \, \partial A_{y,j}}
 - \nonumber \\ -
 \sum\limits_{z,A,t,B} G_{z,A;t,B}
 \frac{\partial^2 \mathcal{F}}{\partial A_{x,i} \, \partial \Phi_{z,A}}
 \frac{\partial^2 \mathcal{F}}{\partial A_{y,j} \, \partial \Phi_{t,B}}  ,
\end{eqnarray}
where $\mathcal{F}$ is the mean-field free energy (\ref{MFFunctional}) calculated in the presence of external gauge field $A_{x,i}$ and $G_{z,A;t,B}$ is the propagator of the Hubbard-Stratonovich field defined by the identity
\begin{eqnarray}
\label{HubbardPropagator}
 \sum\limits_{y,B} G_{x,A;y,B} \frac{\partial^2 \mathcal{F}}{\partial \Phi_{y,B} \partial \Phi_{z,C}}
 =
 \delta_{xz} \delta_{AC} .
\end{eqnarray}

 To proceed, we now restrict our analysis to the continuum theory with the effective single-particle Hamiltonian (\ref{ContinuumDiracHamiltonian}) and the action of the Hubbard-Stratonovich field given by (\ref{HSActionContinuum}). An important property of the Hamiltonian (\ref{ContinuumDiracHamiltonian}) which will significantly simplify the calculations is that the external gauge field $A_{x,i}$ enters it and thus the fermionic part
\begin{eqnarray}
\label{S_def}
 \mathcal{S} = \sum\limits_{\epsilon_i<0} \epsilon_i = \sum\limits_i \frac{\epsilon_i - |\epsilon_i|}{2}
\end{eqnarray}
of the mean-field free energy (\ref{MFFunctional}) in exactly the same way as the component of the Hubbard-Stratonovich field which corresponds to $\Gamma_{A_0} = \tau_3 \otimes \sigma_i = \gamma_0 \gamma_i = \alpha_i$ (up to a trivial factor of $v_F$), and thus we can write
\begin{eqnarray}
\label{DerivativeAreplPhi}
 \frac{\partial^2 \mathcal{F}}{\partial A_{x,i} \, \partial \Phi_{z,A}}
 = v_F \,
 \frac{\partial^2 \mathcal{S}}{\partial \Phi_{x,A_0} \, \partial \Phi_{z,A}} .
\end{eqnarray}
To simplify the notation, we continue to use the symbol of the partial derivative $\partial$ also in the continuum theory, although now it should be understood as a functional derivative.

 It is important to note that the identity (\ref{DerivativeAreplPhi}) holds only for the specific cutoff regularization (\ref{ContinuumDiracHamiltonian}) of the continuum Dirac Hamiltonian without any covariant point-splitting regularization for the covariant derivative with vector gauge field $A_{x,i}$. For lattice regularizations, the gauge field will be associated with lattice links and the Hubbard-Stratonovich field - with lattice sites. Therefore in general they will enter the action in different ways. On the one hand, our cutoff regularization implies the loss of invariance under gauge transformations of $A_{x,i}$. On the other hand, with such a regularization the bare Hamiltonian (\ref{ContinuumDiracHamiltonian}) remains invariant under chiral rotations $h \rightarrow e^{-i \gamma_5 \theta} h e^{i \gamma_5\theta}$. This property is also inherited by the vector current operator $\hat{j}_{x,i}$ which can be obtained as a variation of (\ref{ContinuumDiracHamiltonian}) with respect to $A_{x,i}$. This violation of vector current conservation at the expense of maintaining invariance under chiral rotations is the usual redistribution of the axial anomaly between vector and axial currents \cite{Landsteiner:12:1, Rebhan:10:1, Rubakov:10:1, Buividovich:13:8} and is (sometimes implicitly) used in many derivations of the CME which rely on cutoff regularization, see e.g. \cite{Kharzeev:08:2, Fukushima:10:1, Fukushima:10:2, Kharzeev:09:1, Zyuzin:12:2}. Often the role of the cutoff is played by the Fermi surface at $\epsilon = \mu_A$, which is also the property of a particular regularization \cite{Buividovich:13:8}. It is well known that for free fermions cutoff regularization yields the finite answer for the chiral magnetic conductivity $\sigma_{CME}\lr{\vec{k} \rightarrow 0} = \frac{\mu_A}{2 \pi^2}$ \cite{Kharzeev:08:2, Kharzeev:09:1, Zyuzin:12:2, Buividovich:13:6, Buividovich:13:8}. The conservation of vector current can be always restored by adding a suitable Bardeen counterterm $\delta S \sim \mu_A \int d^4 x \epsilon_{ijk} A_{x,i} \partial_j A_{x,k}$ to the effective action of the theory (which is in our case the mean-field free energy (\ref{MFFunctional})). In this case, the chiral magnetic conductivity should vanish in the limit of zero momentum \cite{Rubakov:10:1}. In this work, we will perform all the calculations in the cutoff regularization, which is much more convenient for our purposes, and then restore the vector current conservation with the help of the Bardeen counterterm. The possibility to use other regularizations for our problem will be discussed in more details in Section~\ref{sec:vacuum_energy}.

 We now use (\ref{DerivativeAreplPhi}) to express the propagator of the Hubbard-Stratonovich field in terms of $\mathcal{S}$ and rewrite the second variation of the mean-field free energy (\ref{mf_response}) as
\begin{eqnarray}
\label{mf_response_simplify1}
 \frac{\delta^2 \mathcal{F}}{\delta A_{x,i} \, \delta A_{y,j}}
 =
 v_F^2 \, \frac{\partial^2 \mathcal{S}}{\partial \Phi_{x,A_0} \, \partial \Phi_{y,B_0}}
 - \nonumber \\ -
 v_F^2 \, \sum\limits_{z,A,t,B}
 \lr{\frac{\partial^2 \mathcal{S}}{\partial \Phi_{z,A} \, \partial \Phi_{t,B}} + \frac{2 \, \delta\lr{z-t} \, \delta_{AB}}{V}}^{-1}
 \times \nonumber \\ \times
 \frac{\partial^2 \mathcal{S}}{\partial \Phi_{x,A_0} \, \partial \Phi_{z,A}}
 \frac{\partial^2 \mathcal{S}}{\partial \Phi_{y,B_0} \, \partial \Phi_{t,B}}  ,
\end{eqnarray}
where $\Gamma_{B_0} = \tau_3 \otimes \sigma_j = \gamma_0 \gamma_j = \alpha_j$, the inverse in the second line is the operator inverse defined as in (\ref{HubbardPropagator}) and all the derivatives should be taken at $\Phi_{x,A} = \Phi^{\star}_{x,A}$ and $A_{x,i}=0$. We have also set $c=1$ and $\Lambda=1$, as in the previous Section. The above expression can be simplified even further by rewriting it in terms of the operators
\begin{eqnarray}
\label{operators_def}
 \Pi_{x,A;y,B} = \frac{\delta^2 \mathcal{F}}{\delta \phi_{x,A} \delta \phi_{y,B}},
 \nonumber \\
 \Sigma_{x,A;y,B} = \left. \frac{\partial^2 \mathcal{S}}{\partial \Phi_{x,A} \partial \Phi_{y,B}} \right|_{\Phi^{\star}} ,
\end{eqnarray}
where $\Pi_{x,A;y,B}$ describes the mean-field response to some external field $\phi_{x,A}$ which enters the Dirac Hamiltonian (\ref{ContinuumDiracHamiltonian}) as $\delta h_{x,y} = \sum\limits_A\phi_{x,A} \Gamma_A \tilde{F}\lr{x-y,\Lambda}$ and $\Sigma_{x,A;y,B}$ is the coordinate representation of the fermionic one-loop correction to the self-energy of the Hubbard-Stratonovich field $\Phi_{x,A}$. The equation (\ref{mf_response_simplify1}) can be then written simply as
\begin{eqnarray}
\label{mf_response_simplify3}
 \frac{\delta^2 \mathcal{F}}{\delta A_{x,i} \, \delta A_{y,j}} = v_F^2 \, \Pi_{x,A_0; y,B_0} .
\end{eqnarray}
After some simple algebraic manipulations, the equation (\ref{mf_response_simplify1}) can be brought into the following form:
\begin{eqnarray}
\label{mf_response_operator}
 \Pi = \Sigma - \Sigma \lr{\Sigma + \frac{2}{V}}^{-1} \Sigma
 = 
 \frac{2}{V} - \frac{4}{V^2} \, \frac{1}{\Sigma + \frac{2}{V}} .
\end{eqnarray}
Using now the identity
\begin{eqnarray}
\label{mf_response_simplify2}
 \Sigma_{x,A;y,B} + \frac{2 \, \delta_{xy} \, \delta_{AB}}{V}
 = \nonumber \\ =
 \frac{\partial^2 \mathcal{F}}{\partial \Phi_{x,A} \partial \Phi_{y,B}} = \lr{G_{x,A;y,B}}^{-1} ,
\end{eqnarray}
where the inverse is again the operator inverse, we can finally rewrite the equation (\ref{mf_response_simplify1}) as
\begin{eqnarray}
\label{mf_response_simplified}
 \frac{\delta^2 \mathcal{F}}{\delta A_{x,i} \, \delta A_{y,j}}
 =
 \frac{2 \, v_F^2 \, \delta\lr{x-y} \delta_{ij}}{V} - \frac{4 \, v_F^2}{V^2} \, G_{x,A_0; y,B_0}  .
\end{eqnarray}

 Let us now discuss the physical meaning of this equation. We see that the Hubbard-Stratonovich field $\Phi_{x,\alpha\beta}$ can mimic any local term in the Dirac Hamiltonian. Thus if the interaction potential $V$ is very large and we can neglect the effect of the quadratic action term of this field (second summand in (\ref{MFFunctional})), any local perturbation $\delta h_{x,y} = \sum\limits_A\phi_{x,A} \Gamma_A \delta_{xy}$ of the Dirac Hamiltonian by some external field $\phi_{x,A}$ can be screened by forming the fermionic condensate $\vev{\hat{\psi}^{\dag}_{x,\alpha} \hat{\psi}_{x,\beta} } \sim \Phi^{\star}_{x,A} \Gamma_{A, \alpha\beta} = \phi_{x,A} \Gamma_{A, \alpha\beta}$. As a result, the free energy does not depend on $\phi_{x,A}$ and the second variation (\ref{mf_response_simplified}) vanishes. From (\ref{mf_response_simplified}) one can see that for large but finite $V$ the response to external perturbations scales as $1/V$. The linear operator which describes such a response consists of a contact term $\frac{2 \delta_{xy} \delta_{ij}}{V}$ and a nontrivial term with the propagator of the Hubbard-Stratonovich field. Remembering that in the strong-coupling regime the fluctuations of the Hubbard-Stratonovich field around the saddle point correspond to the propagation of fermionic bound states, we conclude that this nontrivial term describes particle-hole bound states excited by the external field.

\begin{figure}[htpb]
  \centering
  \includegraphics[width=6cm]{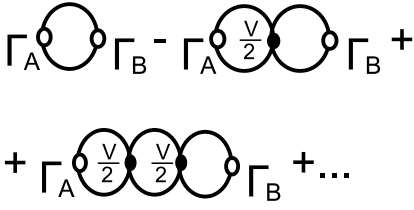}\\
  \caption{Diagrams which contribute to the mean-field linear response operator (\ref{mf_response_simplified}) in the weak-coupling regime.}
  \label{fig:chain_diagrams}
\end{figure}

 In the weak-coupling regime we can expand (\ref{mf_response_operator}) in powers of $V$ to obtain
\begin{eqnarray}
\label{mf_response_weak_coupling}
 \Pi = \Sigma - \Sigma \frac{V}{2} \Sigma + \Sigma \frac{V}{2} \Sigma \frac{V}{2} \Sigma + \ldots  .
\end{eqnarray}
As we will see from what follows, the operator $\Sigma_{x,A; y,B}$ corresponds to the fermionic loop with the insertion of spinor operators $\Gamma_A$ and $\Gamma_B$ at points $x$ and $y$. Then the expansion (\ref{mf_response_weak_coupling}) can be readily interpreted as a sum of an infinite number of chain-like diagrams, as illustrated on Fig.~\ref{fig:chain_diagrams}. At sufficiently large interaction potential $V$ geometric series which describe such a sum diverge, and one has to re-sum them and re-interpret the result in terms of particle-hole bound states.

 The last ingredient which we need for the calculation of the mean-field CME response is the operator $\Sigma_{x,A; y,B} = \frac{\partial^2 \mathcal{S}}{\partial \Phi_{x,A} \, \partial \Phi_{y,B}}$ in explicit form. Since we are interested in the static response functions at zero temperature, the most direct way to arrive at the desired result is to use the standard quantum-mechanical perturbation theory for the effective single-particle Hamiltonian (\ref{ContinuumDiracHamiltonian}) and to expand each of its energy levels to the second order in the Hubbard-Stratonovich field $\Phi_{x,A}$ around $\Phi_{x,A}=\Phi^{\star}_{x,A}$. In the abstract bra-ket notation, the first and the second variations of the energy level $\epsilon_i$ with $\Phi$ read
\begin{eqnarray}
\label{energy_expansion_braket}
 \frac{\partial \epsilon_i}{\partial \Phi_{x,A}} = \bra{\Psi_i} \frac{\partial h}{\partial \Phi_{x,A}} \ket{\Psi_i}
 \nonumber \\
 \frac{\partial^2 \epsilon_i}{\partial \Phi_{x,A} \, \partial \Phi_{x,B}}
 = \nonumber \\ =
 \sum\limits_{j \neq i}
 \frac{\bra{\Psi_i} \frac{\partial h}{\partial \Phi_{x,A}} \ket{\Psi_j} \bra{\Psi_j} \frac{\partial h}{\partial \Phi_{y,B}} \ket{\Psi_i}}{\epsilon_i - \epsilon_j}
 + \nonumber \\ +
 \sum\limits_{j \neq i}
 \frac{\bra{\Psi_i} \frac{\partial h}{\partial \Phi_{y,B}} \ket{\Psi_j} \bra{\Psi_j} \frac{\partial h}{\partial \Phi_{x,A}} \ket{\Psi_i}}{\epsilon_i - \epsilon_j}  ,
\end{eqnarray}
where $\ket{\Psi_i}$ is the eigenstate which corresponds to the energy level $\epsilon_i$ of the single-particle effective Hamiltonian. Correspondingly, the operator $\Sigma_{x,A;y,B}$ can be written as
\begin{eqnarray}
\label{S_2nd_var}
 \Sigma_{x,A;y,B} =
 \sum\limits_i \delta\lr{\epsilon_i}
 \frac{\partial \epsilon_i}{\partial \Phi_{x,A}}
 \frac{\partial \epsilon_i}{\partial \Phi_{y,B}}
 + \nonumber \\ +
 \sum\limits_i \theta\lr{-\epsilon_i} \frac{\partial^2 \epsilon_i}{\partial \Phi_{x,A} \, \partial \Phi_{y,B}}
 = \nonumber \\ =
 \sum\limits_i \delta\lr{\epsilon_i}
 \bra{\Psi_i} \frac{\partial h}{\partial \Phi_{x,A}} \ket{\Psi_i}
 \bra{\Psi_i} \frac{\partial h}{\partial \Phi_{y,B}} \ket{\Psi_i}
 + \nonumber \\ +
 \sum\limits_{i: \epsilon_i<0}
 \sum\limits_{j \neq i}
 \frac{\bra{\Psi_i} \frac{\partial h}{\partial \Phi_{x,A}} \ket{\Psi_j} \bra{\Psi_j} \frac{\partial h}{\partial \Phi_{y,B}} \ket{\Psi_i}}{\epsilon_i - \epsilon_j}
 + \nonumber \\ +
 \sum\limits_{i: \epsilon_i<0}
 \sum\limits_{j \neq i}
 \frac{\bra{\Psi_i} \frac{\partial h}{\partial \Phi_{y,B}} \ket{\Psi_j} \bra{\Psi_j} \frac{\partial h}{\partial \Phi_{x,A}} \ket{\Psi_i}}{\epsilon_i - \epsilon_j} .
\end{eqnarray}
The first summand on the r.h.s. of (\ref{S_2nd_var}) originates from energy levels which cross zero in the presence of external perturbations. The second summand is the usual one-loop fermionic contribution to the self-energy of the Hubbard-Stratonovich field. The summation over $i$ is restricted to occupied energy levels with $\epsilon_i < 0$, while summation over $j$ goes over all energy levels which do not coincide with $\epsilon_i$. For our continuum regularized Dirac Hamiltonian (\ref{ContinuumDiracHamiltonian}), the eigenstates $\ket{\Psi_i}$ are labelled by two discrete indices $s, \sigma = \pm 1$ as well as the momentum $\vec{k}$, and the energies $\epsilon_i$ are given by (\ref{MuAEnergyLevels}). The derivative $\frac{\partial h}{\partial \Phi_{x,A}}$ is given by
\begin{eqnarray}
\label{ContinuumHamiltonianDerivative}
 \frac{\partial h_{x;y}}{\partial \Phi_{z,A}} = \delta\lr{x-z} \tilde{F}\lr{x-y,\Lambda} \Gamma_A .
\end{eqnarray}

 We relegate the details of the calculation of $\Sigma_{x,A;y,B}$ to Appendix~\ref{apdx:S_2nd_der}. Since the exact analytic calculation turned out to be very complicated at nonzero effective mass $m$ and chiral chemical potential $\mu_A$, we have used numerical integration to sum over all states in (\ref{S_2nd_var}). We then perform the Fourier transform of $\Sigma_{x,A;y,B}$ with respect to $x$ and $y$ as in (\ref{CME_Kubo}) and assume that the momentum $\vec{k}$ is parallel to the 3rd coordinate axis, $\vec{k} = k_3 \vec{e}_3$.

 In order to discuss the structure of $\Sigma_{AB}\lr{\vec{k}}$ and its physical implications, let us recall that according to (\ref{mf_response_simplify2}) $\Sigma_{x,A;y,B}$ and hence also $\Sigma_{AB}\lr{\vec{k}}$ differs from the inverse propagator of the Hubbard-Stratonovich field $\lr{G_{x,A;y,B}}^{-1}$ only by a term diagonal in $A$ and $B$. Since in the strong-coupling regime this propagator describes particle-hole bound states (mesons in QCD terminology), we can interpret the appearance of the off-diagonal terms in $\Sigma_{AB}\lr{\vec{k}}$ as the mixing between different bound states. Several states are mixed already at zero chiral chemical potential due to spontaneous breaking of chiral symmetry. First, by virtue of the on-shell equation $\partial_{\mu} j_{\mu}^A = 2 m \, \bar{\psi} \gamma_5 \psi$ there is the mixing between the longitudinal component of the axial current $k_i j_{i}^{A} = k_i \psi^{\dag} \, I \otimes \sigma_i \, \psi = k_i \bar{\psi} \gamma_i \gamma_5 \psi$ and the Nambu-Goldstone mode (``pion'') which corresponds to the operator $\psi^{\dag} \, \tau_2 \otimes I \, \psi = \bar{\psi} \gamma_5 \psi$. Here we have used the ``relativistic'' notation $\bar{\psi} = \psi^{\dag} \gamma_0$ in order to facilitate the identification of the bound states discussed here with meson states in QCD. Another similar on-shell equation, $\partial_{\mu} \lr{\bar{\psi} \lrs{\gamma_{\mu}, \gamma_{\nu}} \psi} = 4 m \, \bar{\psi} \gamma_{\nu} \psi$, implies the mixing between the longitudinal component of the tensor excitations $k_i \bar{\psi} \lrs{\gamma_i, \gamma_j} \psi$ and $k_i \bar{\psi} \lrs{\gamma_i, \gamma_0} \psi$ and the vector current and the charge density, correspondingly.

 Nonzero chiral chemical potential $\mu_A$ explicitly breaks parity and hence induces mixing between parity-odd and parity-even states. First, the states created by the operators $\psi^{\dag} \, \tau_3 \otimes \sigma_i \, \psi = \bar{\psi} \gamma_i \psi$ (vector current fluctuations, or vector mesons in QCD terminology) are mixed with the fluctuations of magnetization which are described by the operators $\psi^{\dag} \, \tau_1 \otimes \sigma_i \, \psi = i \epsilon_{i j k} \bar{\psi} \lrs{\gamma_j, \gamma_k} \psi$. Second, the scalar states which correspond to the operator $\psi^{\dag} \, \tau_1 \otimes I \, \psi = \bar{\psi} \psi$ (fluctuations of the chiral condensate, or $\sigma$-meson in QCD terminology) are mixed with the fluctuations of the axial charge density, described by the operator $\psi^{\dag} \, \tau_3 \otimes I \, \psi = \bar{\psi} \gamma_5 \gamma_0 \psi$.

 Moreover, chiral imbalance induces the mixing between the two transverse polarizations of the vector and pseudo-vector excitations. It is precisely this mixing between the transverse fluctuations of the vector current, $j_{1,2} = \psi^{\dag} \, \tau_3 \otimes \sigma_{1,2} \, \psi = \bar{\psi} \gamma_{1,2} \psi$, which leads to the appearance of the nonzero off-diagonal element $\bra{0} \hat{j}_{1} \hat{j}_{2} \ket{0}$ of the vector current correlator and hence to nonzero chiral magnetic conductivity $\sigma_{CME}$ according to the Kubo formula (\ref{CME_Kubo}). We therefore conclude that in the strong-coupling phase the CME current is saturated by vector-like bound states ($\rho$-mesons in QCD terminology) with mixed transverse polarizations and a small admixture of pseudo-vector states of both transverse polarizations. A similar picture of meson mixing has been recently addressed also in QCD effective models \cite{Andrianov:13:2}. In this work it was also pointed out that the mixing between the transverse components of vector mesons can be encoded in the Chern-Simons term in the effective action for the vector mesons.

 After having calculated $\Sigma_{AB}\lr{\vec{k}}$ numerically, we plug it into the equations (\ref{mf_response_simplify2}) and (\ref{mf_response_simplified}) and calculate the Fourier transforms of the anomalous current-current correlators $\vev{j_1 j_2}\lr{k_3}$. We then use the Kubo formula (\ref{CME_Kubo}) in order to find the chiral magnetic conductivity $\sigma_{CME}\lr{k_3}$. Before presenting our results for $\vev{j_1 j_2}\lr{k_3}$ and $\sigma_{CME}$, let us make several remarks on their interpretation.

 First, the vector current defined by the functional derivative of the free energy over the gauge field as in (\ref{CME_Kubo}) is not conserved for our regularization (\ref{ContinuumDiracHamiltonian}) of the Dirac Hamiltonian, see the discussion after equation (\ref{DerivativeAreplPhi}). On the other hand, our regularization preserves the invariance of the Hamiltonian under chiral rotations. For free fermions such a regularization yields a finite answer $\sigma_{CME} = \frac{\mu_A^{\lr{0}}}{2 \pi^2}$ \cite{Kharzeev:08:2, Kharzeev:09:1}. Conservation of vector current can be restored by adding the Bardeen counterterm $S_B \sim \int d^4 x \epsilon_{ijk} A_{x,i} \partial_j A_{x,k}$ to the bare action, which should lead to vanishing chiral magnetic conductivity in the limit of small momentum \cite{Rubakov:10:1}. The second variation of the Bardeen counterterm over the gauge field $\frac{\delta^2 S_B}{\delta A_i \delta A_j} \sim \epsilon_{i j l} k_l$ has the same form as the anomalous current-current correlator which enters the Kubo formula (\ref{CME_Kubo}) for the chiral magnetic conductivity. Since the chiral magnetic conductivity should vanish in a gauge-invariant regularization \cite{Rubakov:10:1}, we conclude that the coefficient before the Bardeen counterterm should be equal to the chiral magnetic conductivity $\sigma_{CME}\lr{\vec{k} \rightarrow 0}$ calculated in terms of the non-conserved current. Taking into account this form of the Bardeen counterterm, we conclude that in order to calculate the anomalous correlator of conserved currents we simply have to subtract the term $i \sigma_{CME}\lr{k_3 \rightarrow 0} k_3$ from $\vev{j_1 j_2}\lr{k_3}$.  This subtraction clearly leads to the vanishing chiral magnetic conductivity at zero momentum, however, now $\sigma_{CME}\lr{k_3}$ approaches constant in the limit of infinite momentum \cite{Buividovich:13:8}. In what follows, the correlator of conserved vector currents is denoted as $\vev{\tilde{j}_1 \tilde{j}_2}\lr{k_3}$.

\begin{figure*}[htpb]
  \centering
  \includegraphics[width=6cm,angle=-90]{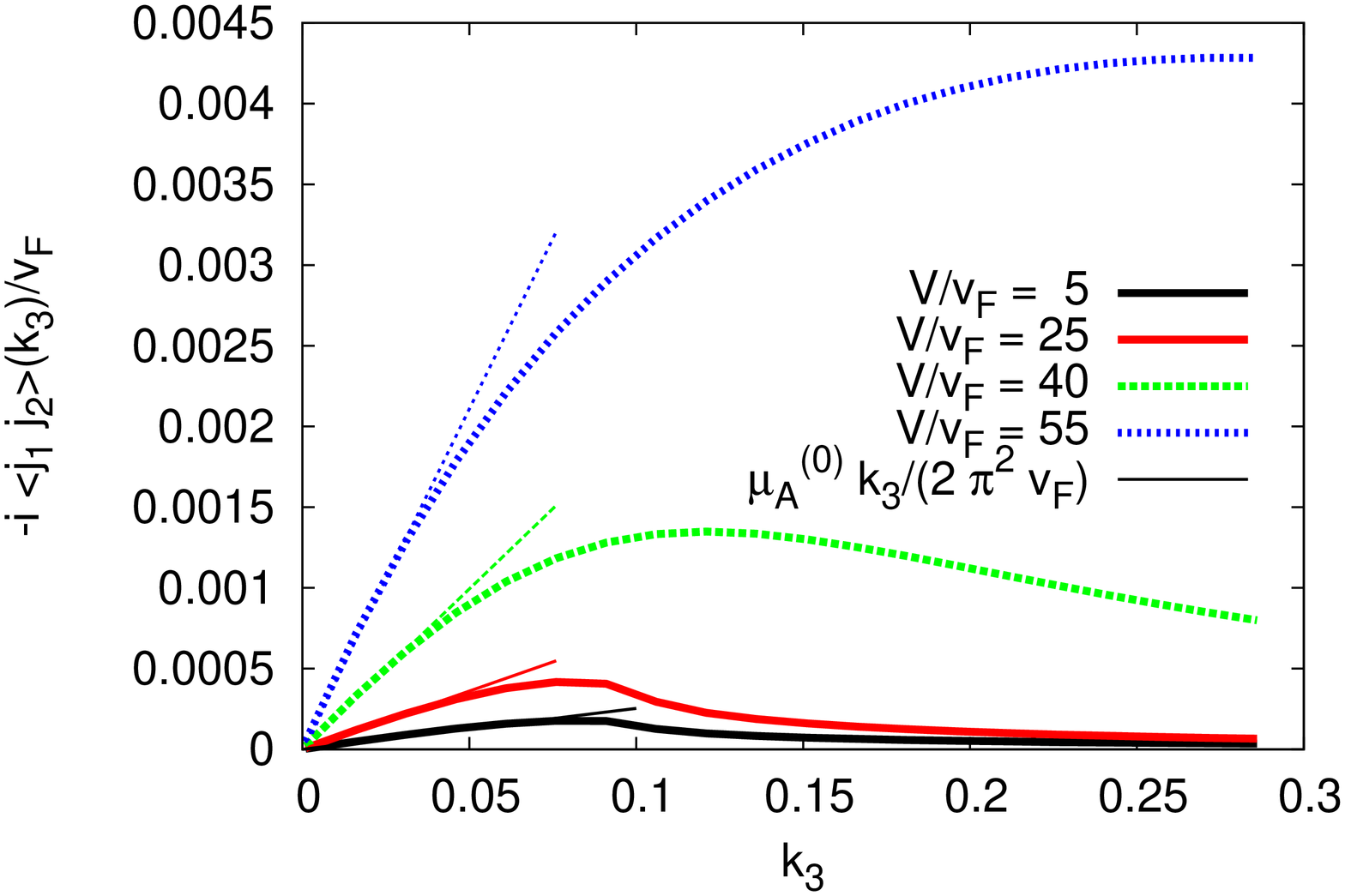}
  \includegraphics[width=6cm,angle=-90]{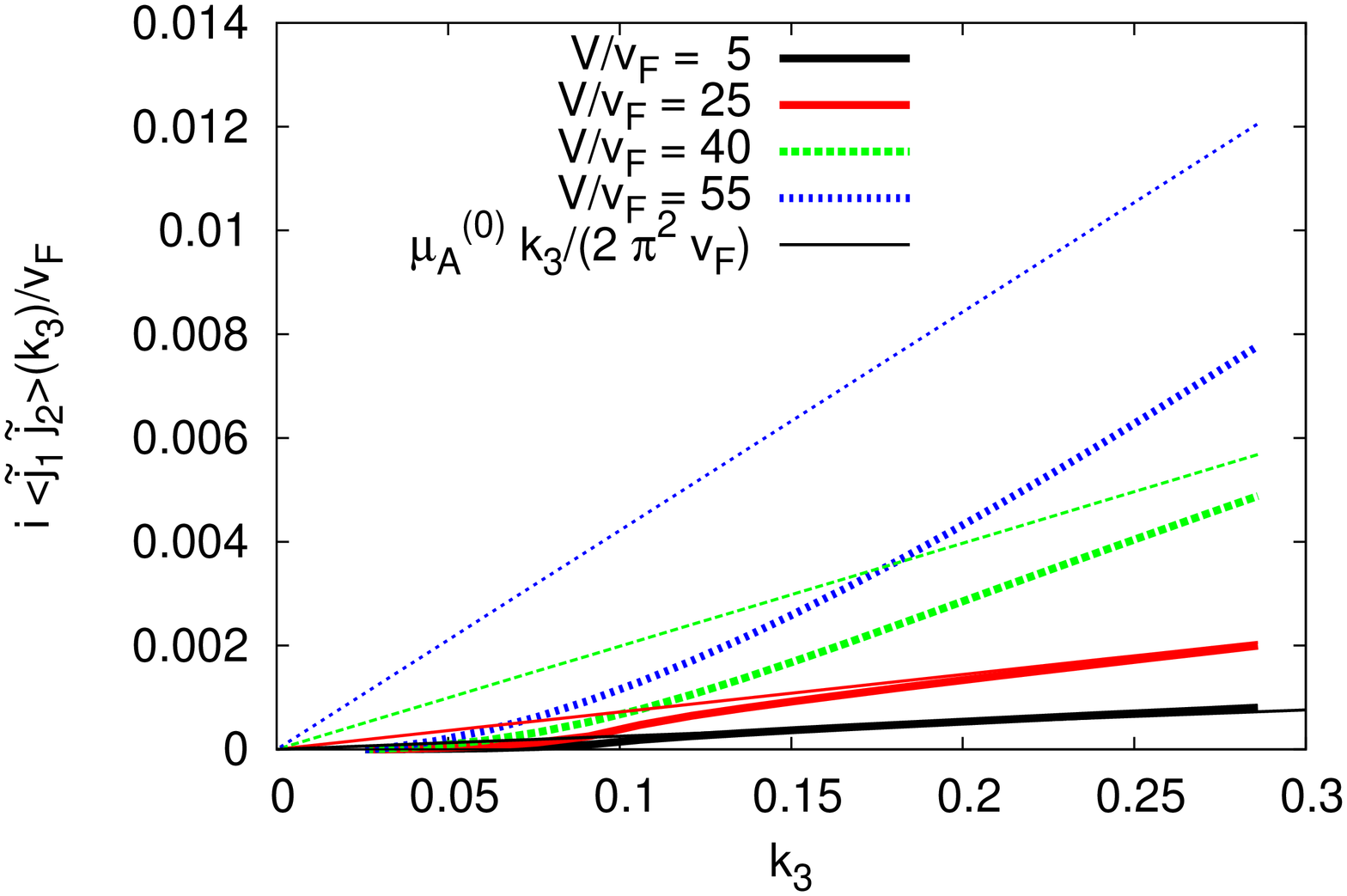}\\
  \includegraphics[width=6cm,angle=-90]{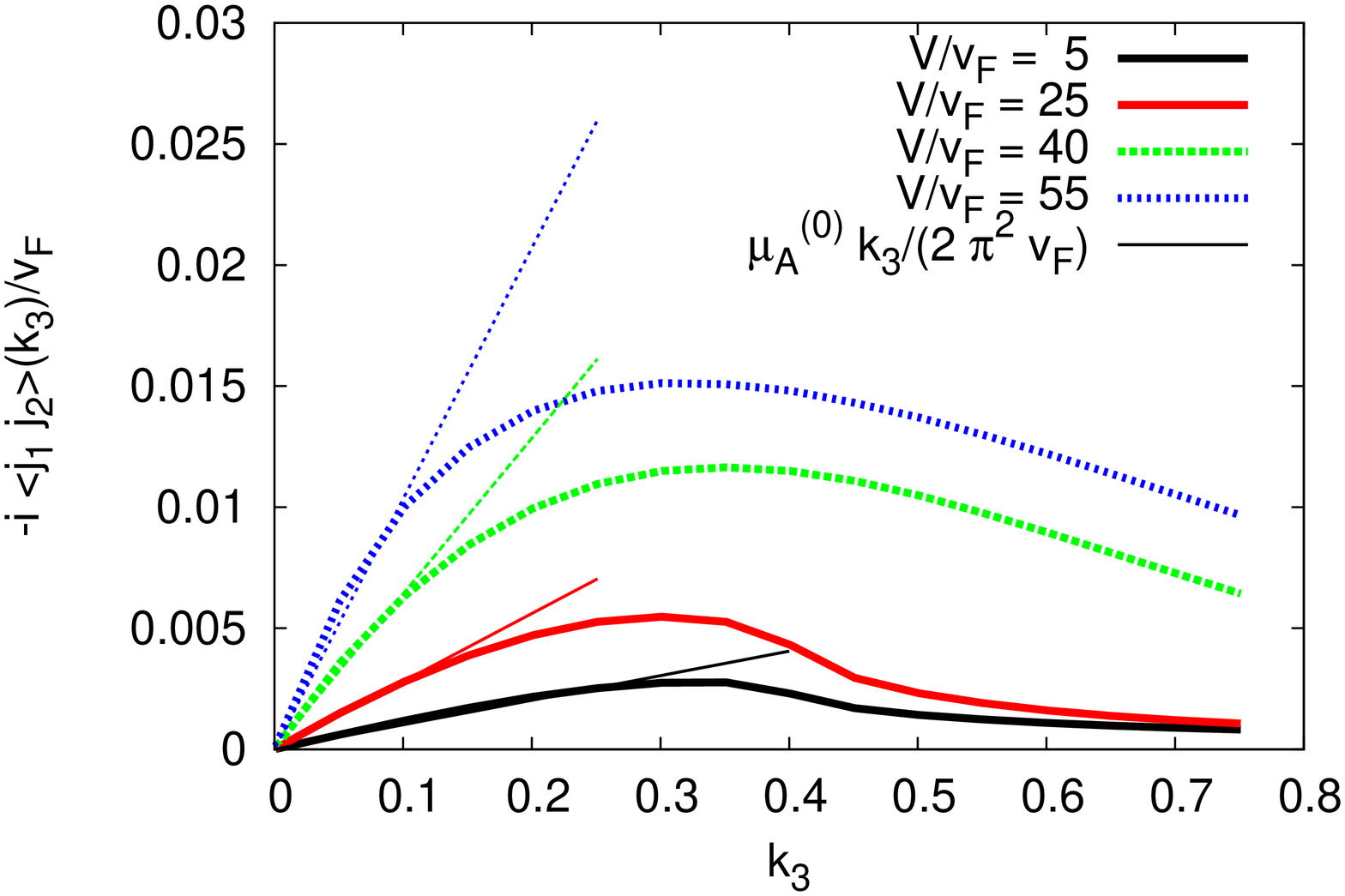}
  \includegraphics[width=6cm,angle=-90]{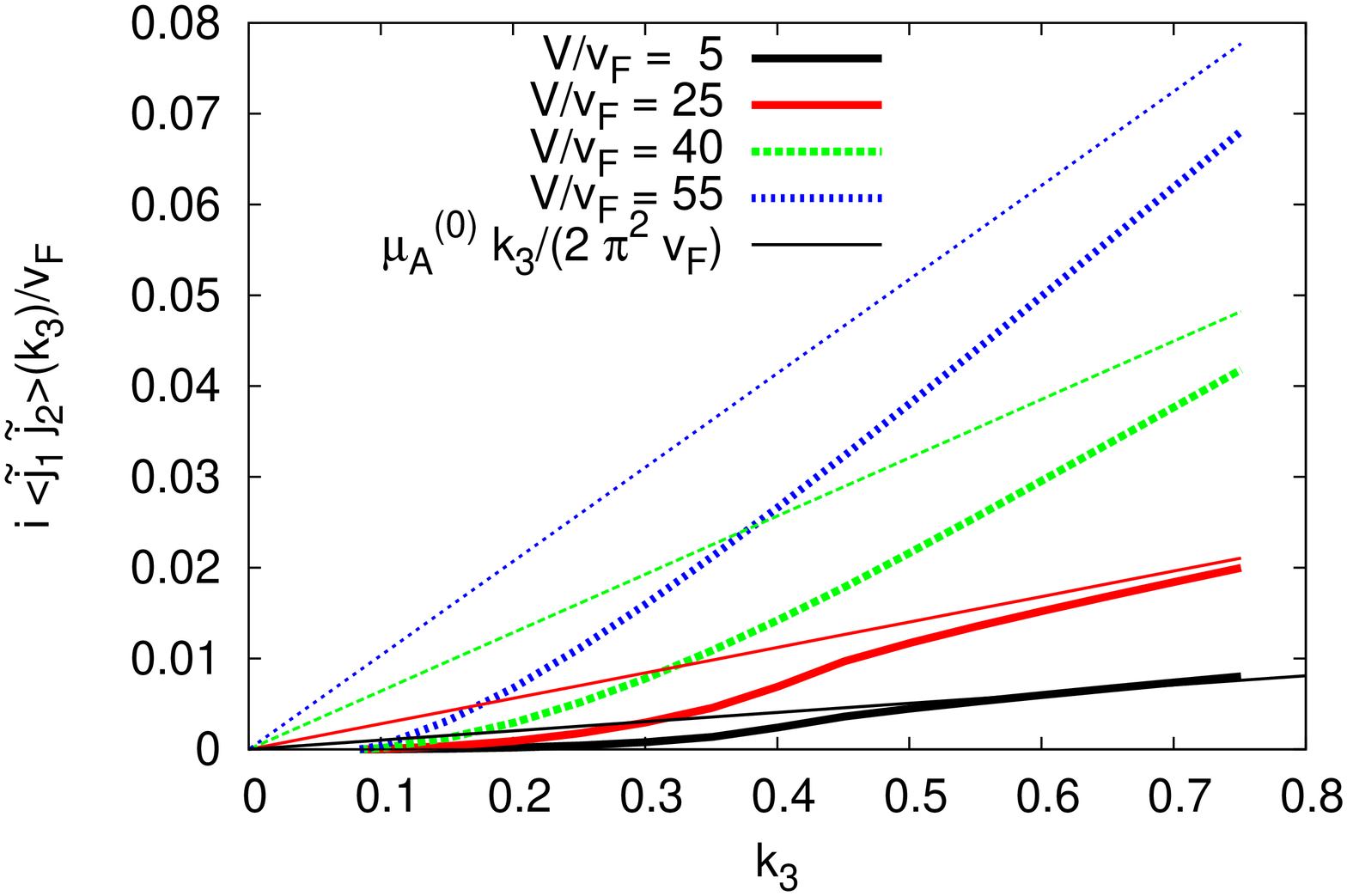}\\
  \caption{The anomalous current-current correlator $\vev{j_1 j_2}\lr{k_3}$ which enters the Kubo relations (\ref{CME_Kubo}) as a function of spatial momentum $k_3$. At the top: at the bare chiral chemical potential $\mu_A^{\lr{0}}/v_F = 0.05$, on the bottom: at $\mu_A^{\lr{0}}/v_F = 0.20$. On the left: current-current correlator calculated in terms of the non-conserved vector currents $j_{1,2}$, on the right: current-current correlator in terms of conserved vector currents $\tilde{j}_{1,2}$ (after the subtraction of the Bardeen counterterm). Thin lines illustrate the asymptotic behavior of $\vev{j_1 j_2}\lr{k_3}$ either in the limit of zero momentum (for non-conserved vector current) or in the limit of large momentum (for conserved current). All numbers are given in units of the UV cutoff scale $\Lambda$.}
  \label{fig:cmc_vs_k}
\end{figure*}

 Second, in order to include the Fermi velocity $v_F < 1$ into our calculations, it is again convenient to express all results in terms of the rescaled variables $\bar{m} = m/v_F$, $\bar{\mu}_A = \mu_A/v_F$ and $\bar{V} = V/v_F$. As discussed in Appendix~\ref{apdx:S_2nd_der}, $\Sigma_{AB}\lr{\vec{k}}$ depends on the Fermi velocity as $\Sigma_{AB}\lr{\vec{k}} \equiv \Sigma_{AB}\lr{\vec{k}; \mu_A, m, v_F} = v_F^{-1} \Sigma_{AB}\lr{\vec{k}; \bar{\mu}_A, \bar{m}, v_F = 1}$. Inserting this relation into (\ref{mf_response_simplify2}) and (\ref{mf_response_simplified}), we see that the ratios $\sigma_{CME}/v_F$ and $\vev{j_1 j_2}/v_F$ depend only on the rescaled variables. Thus in order to calculate the anomalous current-current correlators and the chiral magnetic conductivity at some arbitrary value of $v_F$, one should simply plug the rescaled variables $\bar{m} = m/v_F$, $\bar{\mu}_A = \mu_A/v_F$ and $\bar{V} = V/v_F$ into the result obtained with $v_F = 1$, and finally multiply it by $v_F$. Therefore we give all our numerical results in terms of the rescaled variables introduced above. It is important to note that the momentum variables should not be rescaled. An important consequence of this simple scaling with Fermi velocity is that for free fermions the chiral magnetic conductivity $\sigma_{CME} = v_F \frac{\bar{\mu}_A^{\lr{0}}}{2 \pi^2} = v_F \frac{\mu_A^{\lr{0}}}{2 \pi^2 v_F} = \frac{\mu_A^{\lr{0}}}{2 \pi^2}$ does not depend on $v_F$.

 The results for the anomalous current-current correlators at different values of the interaction potential $V$ and the bare chiral chemical potential $\mu_A^{\lr{0}}$ are presented on Fig.~\ref{fig:cmc_vs_k}. Left and right plots represent the correlators $\vev{j_1 j_2} \lr{k_3}$ and $\vev{\tilde{j}_1 \tilde{j}_2}\lr{k_3}$ of the non-conserved and conserved vector currents, respectively. The plots at the top and on the bottom correspond to bare chiral chemical potential $\mu_A^{\lr{0}}/v_F = 0.05$ and $\mu_A^{\lr{0}}/v_F = 0.20$. One can immediately see that the anomalous current-current correlators $\vev{j_1 j_2} \lr{k_3}$ and $\vev{\tilde{j}_1 \tilde{j}_2}\lr{k_3}$ both grow with $V$ for all values of the momentum $k_3$, both in the perturbative regime at $V < V_c$ and in the strongly coupled regime at $V > V_c$. The relative growth is even more pronounced for the smaller value of the chiral chemical potential. This enhancement of the anomalous current response for interacting fermions is one of the main conclusions of this work.

\begin{figure*}[htpb]
  \centering
  \includegraphics[width=6cm,angle=-90]{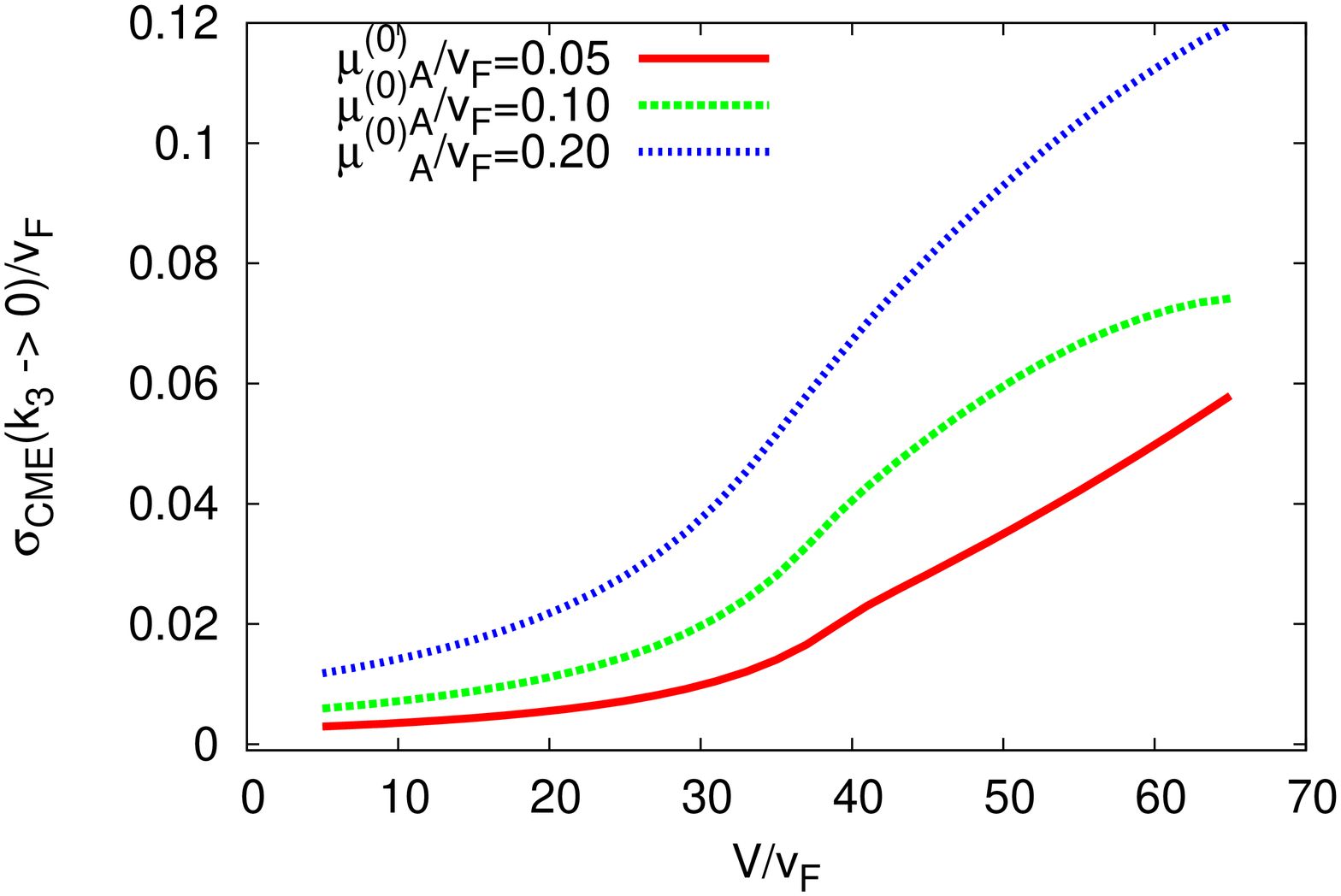}
  \includegraphics[width=6cm,angle=-90]{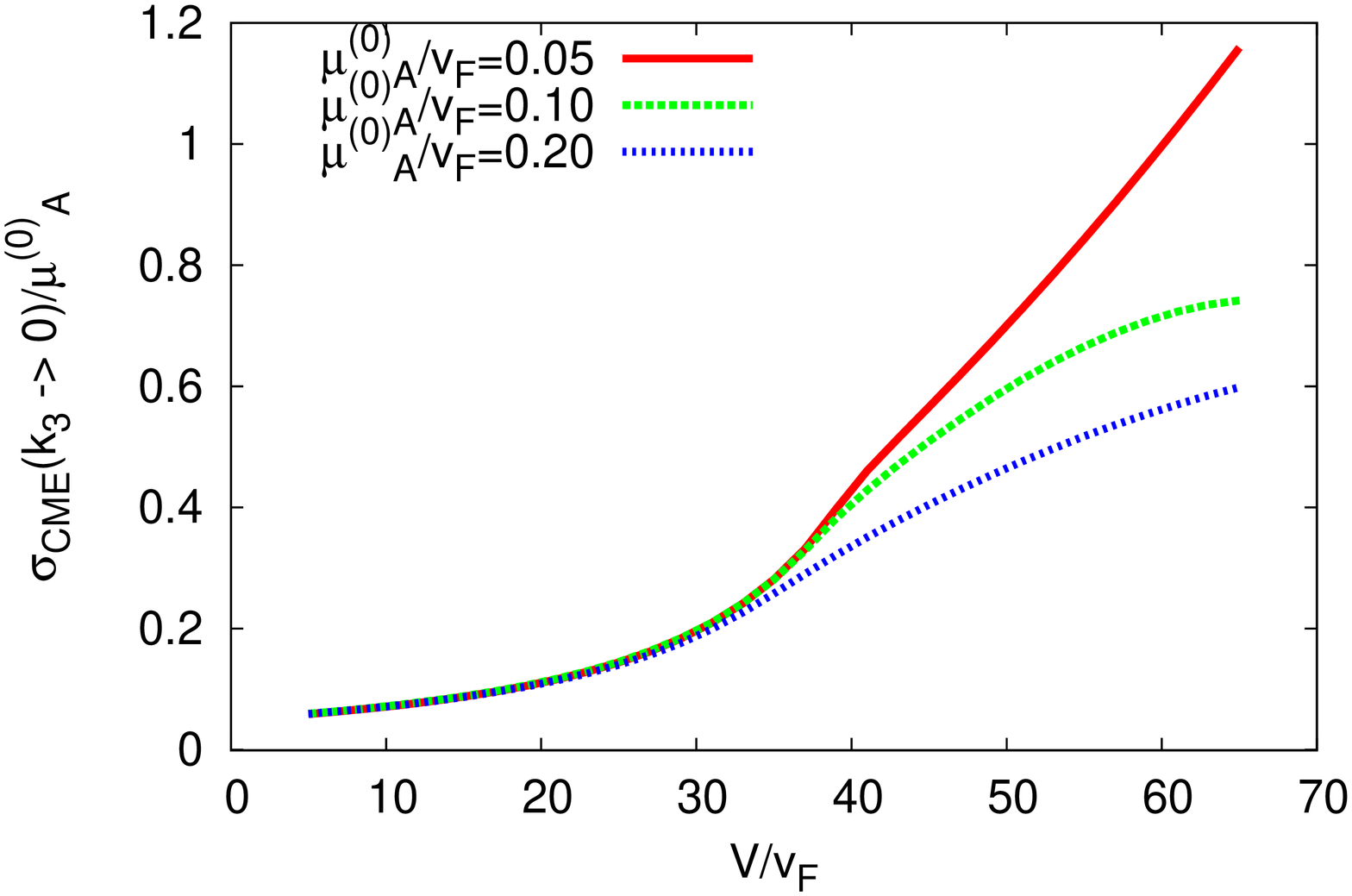}\\
  \caption{Chiral magnetic conductivity $\sigma_{CME}\lr{k_3 \rightarrow 0}$ (calculated in terms of the non-conserved vector current) in the limit of zero momentum as a function of interaction potential $V$ at different values of bare chiral chemical potential $\mu_A^{\lr{0}}$. On the right plot $\sigma_{CME}$ is rescaled by the bare value of the chiral chemical potential $\mu_A^{\lr{0}}$. All numbers are given in units of the UV cutoff scale $\Lambda$.}
  \label{fig:cmc_vs_V}
\end{figure*}

 The slope of the anomalous correlator of non-conserved vector currents at small $k_3$ is the small-momentum limit of the chiral magnetic conductivity $\sigma_{CME}\lr{k_3 \rightarrow 0}$. It is this limit which is relevant for the hydrodynamical description of chirally imbalanced medium \cite{Gynther:10:1, Landsteiner:11:2, Landsteiner:12:1}. $\sigma_{CME}\lr{k_3 \rightarrow 0}$ is plotted on the left plot on Fig.~\ref{fig:cmc_vs_V} as a function of interaction potential $V$ for different values of the bare chiral chemical potential $\mu_A^{\lr{0}}$. In order to illustrate the dependence of $\sigma_{CME}\lr{k_3 \rightarrow 0}$ on $\mu_A^{\lr{0}}$, on the right plot on Fig.~\ref{fig:cmc_vs_V} we also plot the ratios $\sigma_{CME}\lr{k_3 \rightarrow 0}/\mu_A^{\lr{0}}$. Again we see that interactions enhance the chiral magnetic conductivity both in the weak- and in the strong-coupling regimes. At small $V$ $\sigma_{CME}\lr{k_3 \rightarrow 0}$ grows linearly. In the strong-coupling regime this enhancement becomes much stronger. From the right plot on Fig.~\ref{fig:cmc_vs_V} one can again see that the relative increase of the chiral magnetic conductivity is larger for smaller values of $\mu_A^{\lr{0}}$, whereas for larger $\mu_A^{\lr{0}}$ one can note some indications of the saturation of $\sigma_{CME}$ at large $V$.

 Linear growth of $\Sigma_{CME}$ in the perturbative regime can be readily understood from the first terms in the expansion (\ref{mf_response_weak_coupling}). The operator $\Sigma_{AB}\lr{\vec{k}}$ depends on the interaction potential $V$ only through the renormalized mass $m$ and the chiral chemical potential $\mu_A$. Since at small $V$ their values are very close to the bare values $m^{\lr{0}} = 0$ and $\mu_A^{\lr{0}}$ (see Fig.~\ref{fig:mass_vs_V_logscale} and Fig.~\ref{fig:mf_condensates}), in the weak-coupling regime we can neglect the $V$ dependence of $\Sigma_{AB}\lr{\vec{k}}$. In particular, for the off-diagonal part of $\Sigma_{A_0 \, B_0}\lr{\vec{k}}$ which corresponds to the anomalous current-current correlator we can simply use the free-fermion result $\frac{i \mu_A^{\lr{0}} k_3}{2 \pi^2}$. Inserting this expression into (\ref{mf_response_weak_coupling}), (\ref{mf_response_simplify3}) and (\ref{CME_Kubo}) and keeping only the first two terms of the expansion in powers of $V$ as well as only the leading power of $\mu_A$, we obtain
\begin{eqnarray}
\label{cmc_weak_coupling}
 \Pi_{A_0 \, B_0}\lr{\vec{k}} = \Sigma_{A_0 \, B_0}\lr{\vec{k}}
 - \nonumber \\ -
 \Sigma_{A_0 \, A_0}\lr{\vec{k}} \frac{V}{2} \Sigma_{A_0 \, B_0}\lr{\vec{k}}
 - \nonumber \\ -
 \Sigma_{A_0 \, B_0}\lr{\vec{k}} \frac{V}{2} \Sigma_{B_0 \, B_0}\lr{\vec{k}} .
\end{eqnarray}
Direct calculation shows that the diagonal elements $\Sigma_{A_0 \, A_0}\lr{\vec{k}} = \Sigma_{B_0 \, B_0}\lr{\vec{k}}$ are negative, and thus the first perturbative contribution increases $\Pi_{A_0 \, B_0}\lr{\vec{k}}$ and hence by virtue of the relations (\ref{mf_response_simplify3}) and (\ref{CME_Kubo}) also the chiral magnetic conductivity.

 In the strong-coupling regime, the growth of the chiral magnetic conductivity is a nontrivial interplay between the ``screening'' factor $1/V^2$ in the second summand of (\ref{mf_response_simplified}) and the growth of renormalized chiral chemical potential with $V$, which makes the off-diagonal elements $G_{A_0 B_0}$ larger.

\section{Vacuum energy in the presence of chiral chemical potential}
\label{sec:vacuum_energy}

\begin{figure}[htpb]
  \centering
  \includegraphics[width=6cm,angle=-90]{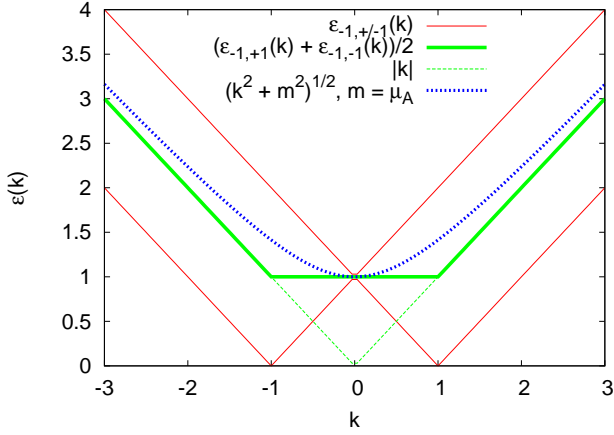}\\
  \caption{Schematic illustration of the dispersion relation in the presence of chiral chemical potential.}
  \label{fig:schematic_dispersion_muA}
\end{figure}

 From the mean-field analysis of Section~\ref{sec:phase_transition} we have seen that interactions tend to increase the chiral chemical potential $\mu_A$. This behavior can be readily explained using the following simple argument. For simplicity let us assume that the Dirac mass is zero, so that the energy levels of the Dirac Hamiltonian $h_0 = -i \alpha_i \nabla_i + \mu_5 \gamma_5$ read
\begin{eqnarray}
\label{massless_energy_levels}
 \varepsilon_{s,\sigma}\lr{\vec{k}} = s ||\vec{k}| - \sigma \mu_A|  ,
\end{eqnarray}
where we have assumed that the Fermi velocity $v_F$ is equal to unity for the sake of brevity.
We now calculate the fermionic contribution to the free energy (\ref{S_def}):
\begin{eqnarray}
\label{massless_fe}
 \mathcal{F} = - \int \frac{d^3\vec{k}}{\lr{2 \pi}^3} \sum\limits_{\sigma = \pm 1} \varepsilon_{-1,\sigma}\lr{\vec{k}} .
\end{eqnarray}
Let us group the summands with the same momentum $\vec{k}$ and different values of $\sigma$:
\begin{eqnarray}
\label{VacuumEnergyWithMuA}
 \varepsilon_{-1,+1}\lr{\vec{k}} + \varepsilon_{-1,1}\lr{\vec{k}}
  = \nonumber \\ =
 -||\vec{k}| - \mu_A| - ||\vec{k}| + \mu_A|
 = 
 \begin{cases}
  -2 \mu_A, & |\vec{k}| < \mu_A \\
  -2 |\vec{k}|, & |\vec{k}| \ge \mu_A \\
 \end{cases}  ,
\end{eqnarray}
After such rearrangement of summands the sum in (\ref{massless_fe}) looks similarly to the free energy of massless Dirac fermions without any chemical potentials, however, now the tip of the Dirac cone is effectively chopped off at the level $\epsilon = \mu_A$ (see Fig.~\ref{fig:schematic_dispersion_muA} for an illustration). This means that effectively all the energy levels of Dirac fermions became lower, and hence the vacuum energy also decreased. Therefore it is energetically advantageous for a system of Dirac fermions to develop nonzero chiral chemical potential, or just to increase its value if the bare value $\mu_A^{\lr{0}}$ is different from zero. However, in the absence of bare chiral chemical potential a more efficient way of lowering the vacuum energy is simply the generation of effective mass, and the system prefers to spontaneously break chiral symmetry. It is interesting to note that although at $\mu_A^{\lr{0}} = 0$ nonzero value of $\mu_A$ cannot be generated, there can be still large fluctuations of $\mu_A$. It was conjectured recently in \cite{Chao:13:1, Yu:14:1} that these fluctuations might be the origin of the so-called inverse magnetic catalysis in QCD.

\begin{figure}[htpb]
  \centering
  \includegraphics[width=6cm,angle=-90]{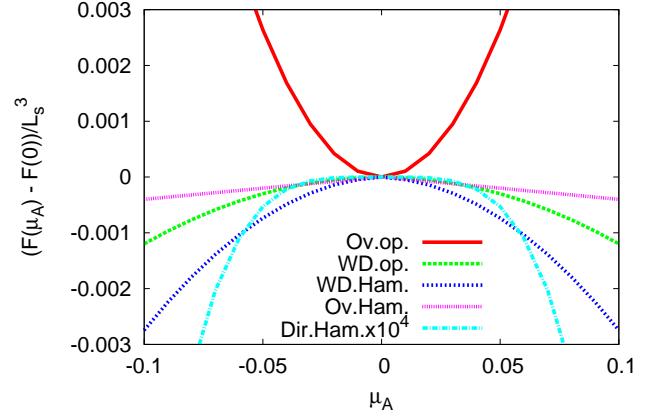}\\
  \caption{Free energy as a function of the chiral chemical potential $\mu_A$ for different regularizations of the Dirac operator/Hamiltonian. ``Ov.op'' is for the overlap Dirac operator, ``WD.op.'' is for the Wilson-Dirac operator with zero bare mass, ``WD.Ham.'' is for the Wilson-Dirac Hamiltonian, ``Ov.Ham'' is for the overlap Hamiltonian and ``Dir.Ham.'' is for the continuum Dirac Hamiltonian (\ref{ContinuumDiracHamiltonian}).}
  \label{fig:fe_vs_muA}
\end{figure}

 However, this simple argument is based on the Hamiltonian with unbounded dispersion relation, for which the vacuum energy is divergent. We therefore have to regularize the problem in some consistent way. In this work our choice was the cutoff regularization (\ref{ContinuumDiracHamiltonian}). However, cutoff regularization breaks gauge invariance, which we then restore by adding the Bardeen counterterm to the action. It is therefore compelling to check whether our results are still valid in more consistent regularizations which automatically preserve gauge invariance (and also preferably the chiral symmetry) of the theory. Here we will check whether the vacuum energy is still lowered by the chiral chemical potential $\mu_A$ for several different regularizations of the continuum Dirac Hamiltonian. We show the plots of the vacuum energy (defined as in (\ref{S_def})) as a function of $\mu_A$ on Fig.~\ref{fig:fe_vs_muA}. The vacuum energy at $\mu_A = 0$ is subtracted from the result. For comparison with different regularizations, we also plot the vacuum energy for our continuum Hamiltonian (\ref{ContinuumDiracHamiltonian}).

 Let us first consider local lattice regularizations, which are natural in the context of condensed matter physics. For instance, a simple model description of Weyl semimetals is provided by the Wilson-Dirac Hamiltonian \cite{Sekine:13:1, Vazifeh:13:1, Hosur:13:1} of the following form:
\begin{eqnarray}
\label{WDHam}
 h_{WD}\lr{\vec{k}, \mu_A} =
 \left(
   \begin{array}{cc}
     \dslash{k} - \mu_A & \Delta\lr{\vec{k}} \\
     \Delta\lr{\vec{k}} & -\dslash{k} + \mu_A \\
   \end{array}
 \right) ,
\end{eqnarray}
where $\dslash{k} = a^{-1} \sum\limits_{i=1}^{3} \sigma_i \sin\lr{a k_i}$, $\Delta\lr{\vec{k}} = 2 a^{-1} \sum\limits_{i=1}^{3} \sin^2\lr{a k_i/2}$ is the Wilson term (for simplicity, we set the Wilson parameter $\rho$ and the Fermi velocity $v_F$ to unity), the momenta $k_i$ belong to the cubic Brillouin zone $k_i \in \lrs{-\pi/a, \pi/a}$ and $a$ is the lattice spacing. Here we have introduced the chiral chemical potential $\mu_A$ by simply adding the term $\mu_A \gamma_5$ to the conventional Wilson-Dirac Hamiltonian. Explicit calculation of the vacuum energy on the $20^3$ lattice shows that it also decreases as $\mu_A$ is increased, see Fig.~\ref{fig:fe_vs_muA}.

 Next it is interesting to consider lattice regularizations with exact chiral symmetry, for instance, overlap fermions. Overlap Dirac Hamiltonian without chemical potentials was introduced in (\cite{Creutz:01:1}):
\begin{eqnarray}
\label{OvHam0}
 h_{ov}^{\lr{0}} = a^{-1} \lr{\gamma_0 + \gamma_0 \gamma_5 \sign\lr{\gamma_5 \gamma_0 \lr{a \, h_{WD} - \rho}}} ,
\end{eqnarray}
where $\rho \in \lrs{0, 2}$. The corresponding axial charge operator which commutes with the Hamiltonian (\ref{OvHam}) reads
\begin{eqnarray}
\label{OvAxialCharge}
 Q_A = \frac{\gamma_5 - \sign\lr{\gamma_5 \gamma_0 \lr{a \, h_{WD} - \rho}}}{2}  .
\end{eqnarray}
We then define the overlap Hamiltonian at finite chiral chemical potential as
\begin{eqnarray}
\label{OvHam}
 h_{ov}\lr{\mu_A} = h_{ov}^{\lr{0}} + \mu_A Q_A
\end{eqnarray}
and after going to momentum space explicitly calculate its energy levels on the $20^3$ lattice. We again observe that the vacuum energy is lowered by the chiral chemical potential.

 It might seem tempting to use the lattice Hamiltonians (\ref{OvHam}) or (\ref{WDHam}) for our study in order to avoid the ambiguities of UV regularization. Let us note, however, that since lattice chiral rotations for the chirally invariant overlap Hamiltonian (\ref{OvHam}) are necessarily non-local \cite{Creutz:01:1}, the on-site four-fermion interaction term in (\ref{HamiltonianInitial}) will anyway break lattice chiral symmetry. Therefore if one would like to study lattice chiral fermions in the Hamiltonian formalism, one should necessarily consider more complicated interactions than in (\ref{HamiltonianInitial}), for example, gauge-mediated interactions. The problem with the Wilson-Dirac Hamiltonian (\ref{WDHam}) is that the chiral symmetry is broken from the very beginning by the Wilson term. Because of that, one cannot really observe the standard picture of chiral symmetry breaking for Wilson-Dirac fermions. Instead, one finds the strongly-coupled phase with broken parity (Aoki phase, \cite{Aoki:84:1}). While a detailed study of the influence of chiral imbalance on this phase is certainly interesting in relation with transport properties of Weyl semimetals and topological insulators, it is beyond the scope of this paper, and we postpone it for further work.

 In the context of relativistic quantum field theory it is more usual to work with the Dirac operator rather than the Dirac Hamiltonian. For some Dirac operator $\mathcal{D}$ on the $L_t \times L_s^3$ lattice we define the vacuum energy as
\begin{eqnarray}
\label{fe_dirac_op}
 \mathcal{S} = -\log{ \det{\mathcal{D}}}/(a L_t) .
\end{eqnarray}
As the temporal lattice extent $L_t \rightarrow \infty$, this definition becomes equivalent to (\ref{S_def}) up to finite-spacing artifacts.

 We continue our comparative study with the Wilson-Dirac operator at finite chiral chemical potential, which was first introduced in \cite{Yamamoto:11:1}:
\begin{eqnarray}
\label{WDOp}
 \mathcal{D}_{WD}\lr{k_0, \vec{k}, \mu_A}
  = \nonumber \\ =
\left(
     \begin{array}{cc}
      \Delta\lr{\vec{k}} + \frac{2}{a} \sin^2\lr{\frac{a k_0}{2}} & \frac{i}{a} \sin\lr{a k_0 - i \mu_A} + \dslash{k} \\
      \frac{i}{a} \sin\lr{a k_0 + i \mu_A} - \dslash{k}  & \Delta\lr{\vec{k}} + \frac{2}{a} \sin^2\lr{\frac{a k_0}{2}} \\
    \end{array}
   \right) .
\end{eqnarray}
A practical advantage of this operator is that the chiral chemical potential does not break time-reversal invariance and thus allows for efficient Monte-Carlo simulations which are free of the sign problem \cite{Yamamoto:11:1}. Explicit calculation of the vacuum energy (\ref{fe_dirac_op}) for this operator on the $20 \times 20^3$ lattice shows that again the chiral chemical potential lowers the vacuum energy, see Fig.~\ref{fig:fe_vs_muA}. This result is not surprising, as the Wilson-Dirac operator is naturally obtained as one goes from Hamiltonian to the path integral formalism for Wilson-Dirac fermions \cite{MontvayMuenster}.

 Now let us consider lattice Dirac operator which preserves exact chiral symmetry. An example is the overlap Dirac operator at finite chiral chemical potential, which was introduced in \cite{Buividovich:13:6}. In contrast to previous findings, it turns out that for this operator the chiral chemical potential increases the vacuum energy (\ref{fe_dirac_op}), see Fig.~\ref{fig:fe_vs_muA}.

 In order to understand somehow this quite unexpected result, let us remember that overlap fermions can be considered as a limiting case of the Pauli-Villars regularization with infinitely many regulator fields \cite{Slavnov:93:1, Neuberger:93:1}. For simplicity, let us consider only one Pauli-Villars regulator for the continuum Dirac operator $\mathcal{D}\lr{\mu_A, m} = \gamma_{\mu} \partial_{\mu} + \gamma_0 \gamma_5 \mu_A + m$. We then have to replace the logarithm of the determinant of the Dirac operator in (\ref{fe_dirac_op}) by the difference $\log{\det{\mathcal{D}\lr{\mu_A, m}}} - \log{\det{\mathcal{D}\lr{\mu_A, M}}}$, where $M$ is the very large mass of the regulator field. Note that there is no consistent many-particle Hamiltonian associated with the regularized action (formally, regulator fields can be thought of as bosons which violated the spin-statistics relation). Calculating now the derivative $\frac{\partial^2 \mathcal{F}}{\partial \mu_A^2}$ at $\mu_A = 0$ for this Pauli-Villars regularization, we arrive at the following result:
\begin{eqnarray}
\label{fe_PV_derivative}
 - \left. \frac{\partial^2 \mathcal{F}}{\partial \mu_A^2} \right|_{\mu_A = 0}
 = \nonumber \\ =
 \int \frac{dk_0 d^3\vec{k}}{4 \pi^4}
 \lr{\frac{m^2 + k_0^2 - \vec{k}^2}{m^2 + k_0^2 + \vec{k}^2}
 -
 \frac{M^2 + k_0^2 - \vec{k}^2}{M^2 + k_0^2 + \vec{k}^2}
 }
 = \nonumber \\ =
 \int \frac{dk k^2}{\pi^2}
 \lr{\frac{m^2}{\lr{m^2 + k^2}^{3/2}}
 -
 \frac{M^2}{\lr{M^2 + k^2}^{3/2}}
 } .
\end{eqnarray}
The integral in the last line is divergent and requires further regularization. However, it is easy to see that the integrand is predominantly negative if $m \ll M$, therefore the sign of $\frac{\partial^2 \mathcal{F}}{\partial \mu_A^2}$ should be positive. This calculation shows that also for the Pauli-Villars regularization the chiral chemical potential increases the vacuum energy. The reason is simply the contribution of the regulator fermions, which also feel the chiral chemical potential.

 Our observations here suggest that there are two interpretations of the chiral chemical potential for a system of Dirac fermions. One is valid for condensed-matter-style models, for which the Dirac sea consists of a finite number of levels with real fermions occupying them. This is the case for our regularization \ref{ContinuumDiracHamiltonian}, the Wilson-Dirac Hamiltonian (\ref{WDHam}), the corresponding Wilson-Dirac operator (\ref{WDOp}) and the overlap Dirac Hamiltonian (\ref{OvHam}). In this case the vacuum energy is always lowered in the presence of chiral imbalance. Another interpretation might be more relevant for relativistic quantum field theories with chiral fermions. In this case one has to subtract the infinite contribution of the Dirac sea in the energy density by using either finite or infinite number of regulator fermions. After this subtraction the contribution of regulator fermions leads to the increase of the vacuum energy in the presence of chiral imbalance. Of course, one can expect that if we switch to this another interpretation of the chiral chemical potential, the results presented in this work might change.

\section{Conclusions and discussion}
\label{sec:conclusions}

 In this work we have reported on a mean-field study of spontaneous breaking of chiral symmetry for Dirac fermions with chiral imbalance and contact interactions between charges. Our main conclusion is the enhancement of the renormalized chiral chemical potential and the chiral magnetic conductivity due to interactions. This enhancement is found both in the weak-coupling perturbative regime and in the strongly-coupled phase with broken chiral symmetry. We believe that this effect is a combination of both perturbative radiative corrections and non-perturbative effects associated with spontaneous breaking of chiral symmetry.

 Perturbative corrections to anomalous transport coefficients are in general allowed if the corresponding currents are coupled to dynamical gauge fields \cite{Miransky:13:1, Jensen:13:1, Gursoy:14:1}, which are in our case mimicked by the contact interactions between electric charges. As discussed in Section~\ref{sec:cme}, our mean-field analysis in fact corresponds to summation of an infinite number of chain-like diagrams. At the level of the single-particle Dirac Hamiltonian another possible source of corrections to the CME current is the renormalization of the Fermi velocity, which is not taken into account in this work.

 Non-perturbative corrections due to spontaneous chiral symmetry breaking can be expected since the appearance of the massless Goldstone modes and the effective mass term in general invalidate the assumptions underlying the proofs of the non-renormalization of anomalous transport coefficients \cite{Son:12:2, Stephanov:12:1, Jensen:12:1, Jensen:12:2, Banerjee:12:1}. Another argument is that the chiral magnetic conductivity can be related to a certain correlator of two vector and one axial currents, which is not renormalized in massless QCD but can get nontrivial non-perturbative corrections in a phase with broken chiral symmetry \cite{Buividovich:13:8, Vainshtein:03:1, Knecht:04:1}. It should be stressed that the enhancement of the chiral magnetic conductivity cannot be explained by the enhancement of the chiral imbalance alone. We also note that a somewhat similar study has been previously reported in \cite{Fukushima:10:1}, where dielectric screening, rather than the enhancement, of the chiral magnetic conductivity was found. We note, however, that the mean-field analysis of \cite{Fukushima:10:1} did not systematically take into account all possible channels of fermion condensation in the presence of external magnetic field. In particular, the renormalization of the chiral chemical potential was not taken into account.

 We also find that even at the smallest values of the bare chiral chemical potential $\mu_A^{\lr{0}}$ the transition to the phase with broken chiral symmetry turns into a soft crossover due to Cooper-type instability at small interaction strength. This crossover becomes softer and softer as $\mu_A^{\lr{0}}$ is increased.

 Let us note that while all our conclusions rely on the assumption of spatially homogeneous fermionic condensates (in the absence of external magnetic fields), holographic studies in the Sakai-Sugimoto model suggest that for sufficiently large bare chiral chemical potential $\mu_A^{\lr{0}}$ the true ground state is not homogeneous \cite{Zamaklar:11:1}. It would be very interesting to find whether the corresponding inhomogeneous ground state also exists in our lattice model, which we leave for the further work. Let us note, however, that direct numerical analysis of the Hessian matrix $\frac{\partial^2 \mathcal{F}}{\partial \Phi_{x, A} \, \partial \Phi_{y, B}}$ of the mean-field free energy (\ref{MFFunctional}) in the vicinity of the homogeneous condensate configuration shows that the homogeneous condensate is at least a local minimum - that is, there are no flat or unstable directions in the space of inhomogeneous condensates $\Phi_{x,A}$ near $\Phi_{x,A} \equiv \Phi_A$ which would allow for immediate decay into a non-homogeneous condensate.

 Our analysis indicates also that in the strongly coupled regime the CME current is saturated by vector-like bound particle-hole bound states (vector mesons) with mixed transverse polarizations. On the other hand, in \cite{Fukushima:12:1, Kalaydzhyan:14:1, Ray:12:1} the CME current in the phase with broken chiral symmetry was derived from the contact terms in the low-energy chiral Lagrangian. It is an interesting question whether the contribution of vector mesons found in our work corresponds somehow to these contact terms in the low-energy effective action or it should be additionally taken into account.

 Our study should be most relevant for condensed matter systems with Dirac or Weyl fermions as low-energy excitations. As discussed in Section~\ref{sec:vacuum_energy}, for relativistic quantum field theories with chiral imbalance there is a certain regularization ambiguity, which might potentially change our conclusions. However, we believe that for instance the mixing of transverse vector mesons in the presence of chiral imbalance should also be present in QCD, irrespective of whether chiral symmetry breaking is enhanced or suppressed by chiral chemical potential.

\begin{acknowledgments}
 The author would like to thank G.~Dunne, T.~Kalaydzhyan, D.~Kharzeev and A.~Sadofyev for interesting discussions which stimulated this work, and to B.~Roy and the anonymous referee for their remarks on Cooper-type instability. This work is supported by the S.~Kowalewskaja award from the Alexander von Humboldt foundation.
\end{acknowledgments}

\appendix

\section{Variation of the mean-field free energy with respect to external fields}
\label{apdx:lrt_mf}

 In the mean-field approximation one replaces the free energy $\mathcal{F}/T = -\log \int \mathcal{D}\Phi_{A} e^{-S\lr{\Phi, \phi}}$ by the value of the action $S\lr{\Phi^{\star}, \phi}$ at the saddle point $\Phi^{\star}_A$ of the corresponding path integral over some bosonic field $\Phi_A$. We have also assumed that the action depends on some external fields $\phi_a$. For the sake of brevity, in this Appendix we use condensed index notation, so that the indices $A$ and $a$ of the field variables include both spatial coordinates and also any internal indices of the fields $\Phi$ and $\phi$. We also assume summation over repeated indices.

 The linear response of the observables $j_a = \frac{\partial S}{\partial \phi_a}$ to the perturbations of the external fields $\phi_a$ is given by the second derivative $\frac{\partial^2 \lr{\mathcal{F}/T}}{\partial \phi_a \, \partial \phi_b}$. In this Appendix we calculate this derivative in the mean-field approximation, replacing $\mathcal{F}/T$ with $S\lr{\Phi^{\star}, \phi}$. Since the saddle-point equation
\begin{eqnarray}
\label{apdx_sp_equation}
 \left. \frac{\partial S\lr{\Phi, \phi}}{\partial \Phi_A} \right|_{\Phi^{\star}} = 0
\end{eqnarray}
in general depends on the external fields $\phi_a$, the saddle-point value of $\Phi_A$ is also a function of $\phi_a$: $\Phi^{\star}_A = \Phi^{\star}_A\lr{\phi}$. Therefore when the free energy $\mathcal{F}/T$ is replaced by the saddle-point action $S\lr{\Phi^{\star}\lr{\phi}, \phi}$, one also has to replace the partial derivatives $\frac{\partial}{\partial \phi_a}$ with the variation
\begin{eqnarray}
\label{apdx_variation_def}
 \frac{\delta}{\delta \phi_a} = \frac{\partial}{\partial \phi_a} + \frac{\partial \Phi_A^{\star}\lr{\phi}}{\partial \phi_a} \frac{\partial}{\partial \Phi_A} .
\end{eqnarray}
Then one can write for the second derivative of the free energy over the external fields:
\begin{eqnarray}
\label{apdx_action_2nd_der}
 \frac{\partial^2 \lr{\mathcal{F}/T}}{\partial \phi_a \, \partial \phi_b} =
 \left. \frac{\delta}{\delta \phi_a} \frac{\delta}{\delta \phi_b} S \right|_{\Phi^{\star}}
 = \nonumber \\ =
 \left( \frac{\partial^2 S}{\partial \phi_a \partial \phi_b}
 +
 \frac{\partial \Phi^{\star}_A}{\partial \phi_a} \frac{\partial^2 S}{\partial \Phi_A \partial \phi_b}
 + \right. \nonumber \\ \left. \left. +
 \frac{\partial \Phi^{\star}_A}{\partial \phi_b} \frac{\partial^2 S}{\partial \Phi_A \partial \phi_a}
 +
 \frac{\partial \Phi^{\star}_A}{\partial \phi_a}
 \frac{\partial \Phi^{\star}_B}{\partial \phi_b}
 \frac{\partial^2 S}{\partial \Phi_A \partial \Phi_B} \right) \right|_{\Phi^{\star}}  ,
\end{eqnarray}
where we have omitted the arguments of the action $S\lr{\Phi, \phi}$ and of the saddle-point field $\Phi^{\star}\lr{\phi}$ for the sake of brevity. Due to the identity $\frac{\partial S\lr{\Phi, \phi}}{\partial \Phi_A}|_{\Phi^{\star}} = 0$, there is no term with the second derivative $\frac{\partial^2 \Phi^{\star}_A}{\partial \phi_a \partial \phi_b}$ in the above equation.

 The derivative of the saddle-point field $\Phi^{\star}_A$ over $\phi$ can be calculated by differentiating the saddle-point equation (\ref{apdx_sp_equation}) over $\phi$:
\begin{eqnarray}
\label{apdx_impl_der1}
 \left. \frac{\partial^2 S}{\partial \Phi_A \partial \phi_a} \right|_{\Phi^{\star}}
 +
 \left. \frac{\partial \Phi^{\star}_B}{\partial \phi_a}
 \frac{\partial^2 S}{\partial \Phi_A \partial \Phi_B} \right|_{\Phi^{\star}} = 0 .
\end{eqnarray}
It is now convenient to define the propagator $G_{AB}$ of the field $\Phi_A$ through the identity
\begin{eqnarray}
\label{apdx_prop_def}
 \left. G_{AB} \frac{\partial^2 S}{\partial \Phi_B \partial \Phi_C} \right|_{\Phi^{\star}} = \delta_{AC} ,
\end{eqnarray}
so that the equation (\ref{apdx_impl_der1}) can be written as
\begin{eqnarray}
\label{apdx_impl_der}
 \left. \frac{\partial \Phi^{\star}_A}{\partial \phi_a} \right|_{\Phi^{\star}} = - G_{A B}
 \left. \frac{\partial^2 S}{\partial \Phi_B \partial \phi_a} \right|_{\Phi^{\star}} .
\end{eqnarray}
Inserting this equation into (\ref{apdx_action_2nd_der}) and using (\ref{apdx_impl_der}) once again, we finally arrive at the desired result:
\begin{eqnarray}
\label{apdx_2nd_var}
 \frac{\partial^2 \lr{\mathcal{F}/T}}{\partial \phi_a \, \partial \phi_b} =
 \left. \frac{\delta}{\delta \phi_a} \frac{\delta}{\delta \phi_b} S \right|_{\Phi^{\star}}
 = \nonumber \\ =
 \left. \frac{\partial^2 S}{\partial \phi_a \partial \phi_b} \right|_{\Phi^{\star}}
 -
 \left. G_{A B}
 \frac{\partial^2 S}{\partial \Phi_A \partial \phi_a}
 \frac{\partial^2 S}{\partial \Phi_B \partial \phi_b} \right|_{\Phi^{\star}} .
\end{eqnarray}

\section{One-loop fermionic contribution to the self-energy of the Hubbard-Stratonovich field}
\label{apdx:S_2nd_der}

 In this Appendix we give the details of the calculation of the operator $\Sigma_{x,A;y,B} = \frac{\partial^2 \mathcal{S}}{\partial \Phi_{x,A} \, \partial \Phi_{y,B}}$, where $\mathcal{S}$ given by (\ref{S_def}) is the fermionic contribution to the mean-field free energy (\ref{MFFunctional}). Our starting point are the expressions (\ref{S_2nd_var}) and (\ref{ContinuumHamiltonianDerivative}). The wave functions $\Psi^{s,\sigma}_x\lr{\vec{k}}$ which correspond to the energy levels $\epsilon_{s,\sigma}\lr{\vec{k}}$ are
\begin{eqnarray}
\label{apdx_MuAWaveFunctions}
 \Psi^{s,\sigma}_x\lr{\vec{k}} = \varphi_{s,\sigma}\lr{\vec{k}} \frac{e^{i \vec{k} \cdot \vec{x}}}{\sqrt{L^3}}  ,
 \nonumber \\
 \varphi_{s,\sigma}\lr{\vec{k}} =
 \left(
   \begin{array}{c}
     \sqrt{\frac{1}{2} + \frac{\sigma v_F |\vec{k}| - \mu_A}{2 \varepsilon_{s,\sigma}\lr{\vec{k}}}} \, \eta_{\sigma}\lr{\vec{k}}  \\
     s \, \sqrt{\frac{1}{2} - \frac{\sigma v_F |\vec{k}| - \mu_A}{2 \varepsilon_{s,\sigma}\lr{\vec{k}}}} \, \eta_{\sigma}\lr{\vec{k}} \\
   \end{array}
 \right) ,
\end{eqnarray}
where $\eta_{\sigma}\lr{\vec{k}}$ is the Weyl spinor which is the normalized eigenstate of the operator $k_i \sigma_i$ with eigenvalue $\sigma |\vec{k}|$: $k_i \sigma_i \eta_{\sigma}\lr{\vec{k}} = \sigma |\vec{k}| \eta_{\sigma}\lr{\vec{k}}$. We note also that the energy levels $\varepsilon_{s,\sigma}\lr{\vec{k}}$ which enter the spinor part $\varphi_{s,\sigma}\lr{\vec{k}}$ of the wave functions (\ref{apdx_MuAWaveFunctions}) are the energy levels (\ref{MuAEnergyLevels}) of the unregularized Dirac Hamiltonian.

 We now insert these expressions into (\ref{S_2nd_var}) and perform the Fourier transform with respect to $x$ and $y$, as in (\ref{CME_Kubo}):
\begin{eqnarray}
\label{apdx_energy_expansion_momentum}
 \Sigma_{AB}\lr{\vec{k}} = \frac{1}{L^3} \sum\limits_{x,y} e^{i \vec{k} \cdot \lr{\vec{x} - \vec{y}}}
 \, \Sigma_{x,A;y,B} .
\end{eqnarray}

 We begin with the calculation of the contribution of the second summand on the r.h.s. of (\ref{S_2nd_var}) to $\Sigma_{AB}\lr{\vec{k}}$, which we denote as $\Sigma_{AB}^{\lr{2}}\lr{\vec{k}}$, and later consider the first contribution (denoted as $\Sigma_{AB}^{\lr{1}}\lr{\vec{k}}$) associated with energy levels crossing zero.

 Explicitly performing the Fourier transform (\ref{apdx_energy_expansion_momentum}), we obtain the following expression for $\Sigma_{AB}^{\lr{2}}\lr{\vec{k}}$:
\begin{widetext}
\begin{eqnarray}
\label{apdx_Sigma2_explicit}
 \Sigma_{AB}^{\lr{2}}\lr{\vec{k}}
 =
 \sum\limits_{s_2,\sigma_1,\sigma_2} \int\frac{d^3 l}{\lr{2 \pi}^3}
 \frac{
  \bar{\varphi}_{s_1,\sigma_1}\lr{\vec{q}} \Gamma_A \varphi_{s_2,\sigma_2}\lr{\vec{p}}
  \bar{\varphi}_{s_2,\sigma_2}\lr{\vec{p}} \Gamma_B \varphi_{s_1,\sigma_1}\lr{\vec{q}}
  F\lr{\vec{p}, \Lambda} F\lr{\vec{q}, \Lambda}
 }{\varepsilon_{s_1,\sigma_1}\lr{\vec{q}} F\lr{\vec{q}, \Lambda} - \varepsilon_{s_2,\sigma_2}\lr{\vec{p}} F\lr{\vec{p}, \Lambda} }
 + \nonumber \\ +
 \sum\limits_{s_2,\sigma_1,\sigma_2} \int\frac{d^3 l}{\lr{2 \pi}^3}
 \frac{
  \bar{\varphi}_{s_1,\sigma_1}\lr{\vec{p}} \Gamma_B \varphi_{s_2,\sigma_2}\lr{\vec{q}}
  \bar{\varphi}_{s_2,\sigma_2}\lr{\vec{q}} \Gamma_A \varphi_{s_1,\sigma_1}\lr{\vec{p}}
  F\lr{\vec{p}, \Lambda} F\lr{\vec{q}, \Lambda}
 }{\varepsilon_{s_1,\sigma_1}\lr{\vec{p}} F\lr{\vec{p}, \Lambda} - \varepsilon_{s_2,\sigma_2}\lr{\vec{q}} F\lr{\vec{q}, \Lambda}} ,
\end{eqnarray}
\end{widetext}
where $\vec{p} = \vec{l} + \vec{k}/2$, $\vec{q} = \vec{l} - \vec{k}/2$ and the ``loop momentum'' variable $\vec{l}$ was introduced in order to satisfy the constraints of momentum conservation $\vec{k} + \vec{q} - \vec{p} = 0$. In order to impose the constraint $\epsilon_i < 0$ in (\ref{S_2nd_var}), we set $s_1 = -1$, thus excluding this index from summations in the above expression.

 Remembering that the regulating factor $F\lr{\vec{p}, \Lambda}$ is equal to one for $|\vec{p}| < \Lambda$ and is very small for $|\vec{p}| > \Lambda$, it is easy to see that due to a specific combination of $F\lr{\vec{p},\Lambda}$ and $F\lr{\vec{q},\Lambda}$ the integrand of (\ref{apdx_Sigma2_explicit}) vanishes if $|\vec{p}| = |\vec{l} + \vec{k}/2| > \Lambda$ or $|\vec{q}| = |\vec{l} - \vec{k}/2| > \Lambda$. Correspondingly, in a region with $|\vec{p}| < \Lambda$ and $|\vec{q}| < \Lambda$ one can simply replace these factors by unity, which significantly simplifies the calculations. In particular, it is convenient to perform an explicit summation over $s_2$ and $\sigma_2$ using the identity
\begin{widetext}
\begin{eqnarray}
\label{apdx_inverse_shifted_h}
 \sum\limits_{s_2, \sigma_2}
 \frac{\varphi_{s_2,\sigma_2}\lr{\vec{p}} \bar{\varphi}_{s_2,\sigma_2}\lr{\vec{p}}}{\varepsilon_{s_1,\sigma_1}\lr{\vec{q}} - \varepsilon_{s_2,\sigma_2}\lr{\vec{p}}}
 = 
 \lr{\varepsilon_{s_1,\sigma_1}\lr{\vec{q}} - h\lr{\vec{p}}}^{-1}
 = \nonumber \\ =
 \sum\limits_{\sigma_2 = \pm 1}
 \frac{1}{\lr{v_F |\vec{q}| - \sigma_1 \mu_A}^2 - \lr{v_F |\vec{p}| - \sigma_2 \mu_A}^2}
 \left(
   \begin{array}{cc}
     \varepsilon_{s_1,\sigma_1}\lr{\vec{q}} + \sigma_2 v_F |\vec{p}| - \mu_A & m \\
     m & \varepsilon_{s_1,\sigma_1}\lr{\vec{q}} - \sigma_2 v_F |\vec{p}| + \mu_A \\
   \end{array}
 \right) \otimes \mathcal{P}_{\sigma_2}\lr{\vec{p}}
,
\end{eqnarray}
where $h\lr{\vec{p}} = v_F \alpha_i p_i + m \gamma_0 + \mu_A \gamma_5$ is the Fourier-transformed effective single-particle Hamiltonian (\ref{ContinuumDiracHamiltonian}) without the regulating factor, the first matrix factor in the last line has chiral indices $L$, $R$ and $\mathcal{P}_{\sigma_2}\lr{\vec{p}} = \eta_{\sigma_2}\lr{\vec{p}} \bar{\eta}_{\sigma_2}\lr{\vec{p}} = \frac{1 + \sigma_2 \sigma_i p_i/|\vec{p}|}{2}$ is the projection operator in the spin space which projects the spin on the direction of momentum $\vec{p}$ with sign $\sigma_2$. Similar identity can be also obtained for the second line of (\ref{apdx_Sigma2_explicit}) upon the replacement $\vec{p} \leftrightarrow \vec{q}$. Now it is convenient to completely factor out the chiral and the spin indices into direct products and to rewrite the equation (\ref{apdx_Sigma2_explicit}) as
\begin{eqnarray}
\label{apdx_Sigma2_dirprods}
  \Sigma_{AB}^{\lr{2}}\lr{\vec{k}}
 =
 \sum\limits_{\sigma_1,\sigma_2} \int\limits_{|\vec{p}|,|\vec{q}|<\Lambda}\frac{d^3 l}{\lr{2 \pi}^3} \,
 \frac{\tr\lr{\mathcal{P}_q \sigma_A \mathcal{P}_p \sigma_B}}{r_p^2 - r_q^2}
 \times \nonumber \\ \times
 \left(
  \bar{\varphi}_p \,
  \tau_B \,
  \left(
    \begin{array}{cc}
      r_q - \epsilon_p & m \\
      m & -r_q - \epsilon_p \\
    \end{array}
  \right) \,
  \tau_A \,
  \varphi_p
  -
  \bar{\varphi}_q \,
  \tau_A \,
  \left(
    \begin{array}{cc}
      r_p - \epsilon_q & m \\
      m & -r_p - \epsilon_q \\
    \end{array}
  \right) \,
  \tau_B \,
  \varphi_q
 \right)  ,
\end{eqnarray}
\end{widetext}
where we have introduced the following short-hand notations in order to make the expressions more compact:
\begin{eqnarray}
\label{apdx_pq_notation1}
 r_p = \sigma_1 v_F |\vec{p}| - \mu_A, \quad
 r_q = \sigma_2 v_F |\vec{q}| - \mu_A, \nonumber \\
 \mathcal{P}_p = \frac{1 + \sigma_1 \sigma_i p_i/|\vec{p}|}{2}, \quad
 \mathcal{P}_q = \frac{1 + \sigma_2 \sigma_i q_i/|\vec{q}|}{2}, \nonumber \\
 \varepsilon_p = |\varepsilon_{s_1,\sigma_1}\lr{\vec{p}}| = \sqrt{r_p^2 + m^2}, \nonumber \\
 \varepsilon_q = |\varepsilon_{s_1,\sigma_2}\lr{\vec{q}}| = \sqrt{r_q^2 + m^2}, \nonumber
\end{eqnarray}
\begin{eqnarray}
\label{apdx_pq_notation2}
 \varphi_p = \left(
   \begin{array}{c}
     \sqrt{\frac{1}{2} + \frac{\sigma_1 v_F |\vec{p}| - \mu_A}{2 \varepsilon_{s_1,\sigma_1}\lr{\vec{p}}}} \\
     s_1 \, \sqrt{\frac{1}{2} - \frac{\sigma_1 v_F |\vec{p}| - \mu_A}{2 \varepsilon_{s_1,\sigma_1}\lr{\vec{p}}}} \\
   \end{array}
 \right)
 =
 \left(
   \begin{array}{c}
     \sqrt{\frac{1}{2} - \frac{r_p}{2 \varepsilon_p}} \\
   - \sqrt{\frac{1}{2} + \frac{r_p}{2 \varepsilon_p}} \\
   \end{array}
 \right),  \nonumber \\
 \varphi_q = \left(
   \begin{array}{c}
     \sqrt{\frac{1}{2} + \frac{\sigma_2 v_F |\vec{q}| - \mu_A}{2 \varepsilon_{s_1,\sigma_2}\lr{\vec{q}}}} \\
   s_1 \, \sqrt{\frac{1}{2} - \frac{\sigma_2 v_F |\vec{q}| - \mu_A}{2 \varepsilon_{s_1,\sigma_2}\lr{\vec{q}}}} \\
   \end{array}
 \right)
 =
 \left(
   \begin{array}{c}
     \sqrt{\frac{1}{2} - \frac{r_q}{2 \varepsilon_q}} \\
   - \sqrt{\frac{1}{2} + \frac{r_q}{2 \varepsilon_q}} \\
   \end{array}
 \right)  . \nonumber
\end{eqnarray}
When defining the energies $\varepsilon_p$, $\varepsilon_q$ and the chiral wave functions $\varphi_p$ and $\varphi_q$ we have explicitly taken into account that $s_1 = -1$. By expressing the integrand of (\ref{apdx_Sigma2_dirprods}) in terms of the rescaled variables $\bar{\mu}_A = \mu_A/v_F$, $\bar{m} = m/v_F$ (see Eq.~(\ref{MFFunctionalRescaled}) in Section~\ref{sec:phase_transition}) and correspondingly $\bar{r}_{p,q} = r_{p,q}/v_F$, $\bar{\varepsilon}_{p,q} = \varepsilon_{p,q}/v_F$ one can completely eliminate the Fermi velocity in the integrand of (\ref{apdx_Sigma2_dirprods}). It only appears as an overall factor of $v_F^{-1}$ in front of (\ref{apdx_Sigma2_dirprods}). Thus in order to calculate $\Sigma_{AB}\lr{\vec{k}}$ at some $v_F < 1$, one should simply substitute the rescaled variables $\bar{\mu}_A$ and $\bar{m}$ into the expression obtained with $v_F = 1$, and multiply the result by $v_F^{-1}$.

 The above representation of $\Sigma^{\lr{2}}_{AB}\lr{\vec{k}}$ is especially convenient for both analytic transformations and numerical calculations. First, we see that the traces over the ``chiral'' and the ``spin'' indices completely factorize in the integrand of (\ref{apdx_Sigma2_dirprods}). Explicit calculation of the ``chiral'' part (the trace which involves $\tau_A$ and $\tau_B$) shows that the integrand is only nonzero if $\tau_A = \tau_B$ (diagonal terms) or if $\tau_A = I$, $\tau_B = \tau_2$ and vice versa or $\tau_A = \tau_1$ and $\tau_B = \tau_3$ and vice versa (off-diagonal terms). Assuming that the momentum $\vec{k}$ is parallel to the $3$rd coordinate axis, $\vec{k} = k_3 \vec{e}_3$, one can also see that the ``spin'' part is only nonzero if $\sigma_A = \sigma_B$ (diagonal terms) or if $\sigma_A = I$, $\sigma_B = \sigma_3$ and vice versa or if $\sigma_A = \sigma_1$, $\sigma_B = \sigma_2$ and vice versa. Some of the off-diagonal terms become zero only after integrating out the loop momentum. The appearance of these off-diagonal terms both in the ``chiral'' and the ''spin'' parts of the self-energy of the Hubbard-Stratonovich field implies an interesting picture of mixing between different particle-hole bound states (``mesons'' in the language of QCD) of different spin/parity and polarizations in the presence of parity-breaking chiral chemical potential $\mu_A$. We discuss this picture in more details in the Section~\ref{sec:cme} of the main text.

 After explicitly calculating the integrand of (\ref{apdx_Sigma2_dirprods}), we rewrite the integral over the loop momentum $l$ in the cylindrical coordinates $\vec{l} = \lrc{l_{\perp} \cos\lr{\vartheta}, l_{\perp} \sin\lr{\vartheta}, l_3}$. It is important that the integration region specified by $|\vec{p}| < \Lambda$, $|\vec{q}| < \Lambda$ is also cylindrically symmetric around the $3$rd coordinate axis. Integration over $\vartheta$ can be then performed analytically and additionally removes many terms in the integrand. We are then left with the integral over $l_{\perp}$ and $l_3$, which can be performed analytically at zero mass \cite{Buividovich:13:8}. At nonzero mass, this is no longer possible and we use the \texttt{Cubature} \texttt{C} package \cite{Johnson:Cubature} to calculate the integral.

 Let us now turn to the second contribution to $\Sigma_{x,A;y,B}$, which is associated with the energy levels which cross zero in external field. This contribution comes from the first summand on the r.h.s. of (\ref{S_2nd_var}). We denote it as $\Sigma^{\lr{1}}_{x,A;y,B}$. First we note that obviously this contribution is only nonzero if the effective mass $m$ is zero and there is no gap in the spectrum. If the spectrum has finite gap, infinitely small perturbations cannot move any energy level down to zero. Thus we should consider only the phase with unbroken chiral symmetry. Inserting the explicit form of the eigenstates (\ref{apdx_MuAWaveFunctions}) into (\ref{S_2nd_var}) and performing the Fourier transform (\ref{apdx_energy_expansion_momentum}), one can immediately see that $\Sigma^{\lr{1}}$ contribution is proportional to the delta-function at zero momentum: $\Sigma^{\lr{1}}\lr{\vec{k}} \sim \delta\lr{\vec{k}}$. Moreover, it is only nonzero for $\sigma_{A} = \sigma_{B}$ and $\tau_A = I, \tau_1$ and $\tau_B = I, \tau_1$. This delta-function singularity will result in the usual ``ballistic'' contribution to the transport coefficients and will not affect the anomalous transport. Therefore we disregard it in this work.


\begin{thebibliography}{66}
\expandafter\ifx\csname natexlab\endcsname\relax\def\natexlab#1{#1}\fi
\expandafter\ifx\csname bibnamefont\endcsname\relax
  \def\bibnamefont#1{#1}\fi
\expandafter\ifx\csname bibfnamefont\endcsname\relax
  \def\bibfnamefont#1{#1}\fi
\expandafter\ifx\csname citenamefont\endcsname\relax
  \def\citenamefont#1{#1}\fi
\expandafter\ifx\csname url\endcsname\relax
  \def\url#1{\texttt{#1}}\fi
\expandafter\ifx\csname urlprefix\endcsname\relax\def\urlprefix{URL }\fi
\providecommand{\bibinfo}[2]{#2}
\providecommand{\eprint}[2][]{\url{#2}}

\bibitem[{\citenamefont{Kharzeev et~al.}(2012)\citenamefont{Kharzeev,
  Landsteiner, Schmitt, and Yee}}]{Kharzeev:12:1}
\bibinfo{author}{\bibfnamefont{D.~E.} \bibnamefont{Kharzeev}},
  \bibinfo{author}{\bibfnamefont{K.}~\bibnamefont{Landsteiner}},
  \bibinfo{author}{\bibfnamefont{A.}~\bibnamefont{Schmitt}}, \bibnamefont{and}
  \bibinfo{author}{\bibfnamefont{H.}~\bibnamefont{Yee}},
  \emph{Strongly interacting matter in magnetic fields: an
  overview}, \bibinfo{howpublished}{in Lect. Notes Phys. {Strongly interacting
  matter in magnetic fields} (Springer)} (\bibinfo{year}{2012}),
  \href{http://arxiv.org/abs/1211.6245}{ArXiv:1211.6245}.

\bibitem[{\citenamefont{Fukushima et~al.}(2008)\citenamefont{Fukushima,
  Kharzeev, and Warringa}}]{Kharzeev:08:2}
\bibinfo{author}{\bibfnamefont{K.}~\bibnamefont{Fukushima}},
  \bibinfo{author}{\bibfnamefont{D.~E.} \bibnamefont{Kharzeev}},
  \bibnamefont{and} \bibinfo{author}{\bibfnamefont{H.~J.}
  \bibnamefont{Warringa}}, \bibinfo{journal}{Phys.Rev.D}
  \textbf{\bibinfo{volume}{78}}, \bibinfo{pages}{074033}
  (\bibinfo{year}{2008}), \href{http://arxiv.org/abs/0808.3382}{ArXiv:0808.3382}.

\bibitem[{\citenamefont{Kharzeev et~al.}(2008)\citenamefont{Kharzeev, McLerran,
  and Warringa}}]{Kharzeev:08:1}
\bibinfo{author}{\bibfnamefont{D.~E.} \bibnamefont{Kharzeev}},
  \bibinfo{author}{\bibfnamefont{L.~D.} \bibnamefont{McLerran}},
  \bibnamefont{and} \bibinfo{author}{\bibfnamefont{H.~J.}
  \bibnamefont{Warringa}}, \bibinfo{journal}{Nucl. Phys. A}
  \textbf{\bibinfo{volume}{803}}, \bibinfo{pages}{227} (\bibinfo{year}{2008}),
  \href{http://arxiv.org/abs/0711.0950}{ArXiv:0711.0950}.

\bibitem[{\citenamefont{Volovik}(2003)}]{VolovikHeliumDroplet}
\bibinfo{author}{\bibfnamefont{G.~E.} \bibnamefont{Volovik}},
  \emph{The Universe in a Helium Droplet}
  (\bibinfo{publisher}{Clarendon Press}, \bibinfo{year}{2003}).

\bibitem[{\citenamefont{Zyuzin et~al.}(2012)\citenamefont{Zyuzin, Wu, and
  Burkov}}]{Zyuzin:12:1}
\bibinfo{author}{\bibfnamefont{A.~A.} \bibnamefont{Zyuzin}},
  \bibinfo{author}{\bibfnamefont{S.}~\bibnamefont{Wu}}, \bibnamefont{and}
  \bibinfo{author}{\bibfnamefont{A.~A.} \bibnamefont{Burkov}},
  \bibinfo{journal}{Phys.Rev.B} \textbf{\bibinfo{volume}{85}},
  \bibinfo{pages}{165110} (\bibinfo{year}{2012}), \href{http://arxiv.org/abs/1201.3624}{ArXiv:1201.3624}.

\bibitem[{\citenamefont{Chen et~al.}(2013)\citenamefont{Chen, Wu, and
  Burkov}}]{Burkov:13:1}
\bibinfo{author}{\bibfnamefont{Y.}~\bibnamefont{Chen}},
  \bibinfo{author}{\bibfnamefont{S.}~\bibnamefont{Wu}}, \bibnamefont{and}
  \bibinfo{author}{\bibfnamefont{A.~A.} \bibnamefont{Burkov}},
  \bibinfo{journal}{Phys.Rev.B} \textbf{\bibinfo{volume}{88}},
  \bibinfo{pages}{125105} (\bibinfo{year}{2013}), \href{http://arxiv.org/abs/1306.5344}{ArXiv:1306.5344}.

\bibitem[{\citenamefont{Goswami and Tewari}(2013{\natexlab{a}})}]{Goswami:13:1}
\bibinfo{author}{\bibfnamefont{P.}~\bibnamefont{Goswami}} \bibnamefont{and}
  \bibinfo{author}{\bibfnamefont{S.}~\bibnamefont{Tewari}},
  \emph{Chiral magnetic effect of {W}eyl fermions and its
  applications to cubic noncentrosymmetric metals}
  (\bibinfo{year}{2013}{\natexlab{a}}), \href{http://arxiv.org/abs/1311.1506}{ArXiv:1311.1506}.

\bibitem[{\citenamefont{Goswami and Tewari}(2013{\natexlab{b}})}]{Goswami:12:1}
\bibinfo{author}{\bibfnamefont{P.}~\bibnamefont{Goswami}} \bibnamefont{and}
  \bibinfo{author}{\bibfnamefont{S.}~\bibnamefont{Tewari}},
  \bibinfo{journal}{Phys.Rev.B} \textbf{\bibinfo{volume}{88}},
  \bibinfo{pages}{245107} (\bibinfo{year}{2013}{\natexlab{b}}),
  \href{http://arxiv.org/abs/1210.6352}{ArXiv:1210.6352}.

\bibitem[{\citenamefont{Vazifeh and Franz}(2013)}]{Vazifeh:13:1}
\bibinfo{author}{\bibfnamefont{M.~M.} \bibnamefont{Vazifeh}} \bibnamefont{and}
  \bibinfo{author}{\bibfnamefont{M.}~\bibnamefont{Franz}},
  \bibinfo{journal}{Phys.Rev.Lett.} \textbf{\bibinfo{volume}{111}},
  \bibinfo{pages}{027201} (\bibinfo{year}{2013}), \href{http://arxiv.org/abs/1303.5784}{ArXiv:1303.5784}.

\bibitem[{\citenamefont{Zhou et~al.}(2013)\citenamefont{Zhou, Jiang, Niu, and
  Shi}}]{Zhou:13:1}
\bibinfo{author}{\bibfnamefont{J.}~\bibnamefont{Zhou}},
  \bibinfo{author}{\bibfnamefont{H.}~\bibnamefont{Jiang}},
  \bibinfo{author}{\bibfnamefont{Q.}~\bibnamefont{Niu}}, \bibnamefont{and}
  \bibinfo{author}{\bibfnamefont{J.}~\bibnamefont{Shi}},
  \bibinfo{journal}{Chin.Phys.Lett.} \textbf{\bibinfo{volume}{30}},
  \bibinfo{pages}{027101} (\bibinfo{year}{2013}), \href{http://arxiv.org/abs/1211.0772}{ArXiv:1211.0772}.

\bibitem[{\citenamefont{Wan et~al.}(2011)\citenamefont{Wan, Turner, Vishwanath,
  and Savrasov}}]{Wan:11:1}
\bibinfo{author}{\bibfnamefont{X.}~\bibnamefont{Wan}},
  \bibinfo{author}{\bibfnamefont{A.~M.} \bibnamefont{Turner}},
  \bibinfo{author}{\bibfnamefont{A.}~\bibnamefont{Vishwanath}},
  \bibnamefont{and} \bibinfo{author}{\bibfnamefont{S.~Y.}
  \bibnamefont{Savrasov}}, \bibinfo{journal}{Phys.Rev.B}
  \textbf{\bibinfo{volume}{83}}, \bibinfo{pages}{205101}
  (\bibinfo{year}{2011}), \href{http://arxiv.org/abs/1007.0016}{ArXiv:1007.0016}.

\bibitem[{\citenamefont{Burkov and Balents}(2011)}]{Burkov:11:1}
\bibinfo{author}{\bibfnamefont{A.~A.} \bibnamefont{Burkov}} \bibnamefont{and}
  \bibinfo{author}{\bibfnamefont{L.}~\bibnamefont{Balents}},
  \bibinfo{journal}{Phys.Rev.Lett.} \textbf{\bibinfo{volume}{107}},
  \bibinfo{pages}{127205} (\bibinfo{year}{2011}), \href{http://arxiv.org/abs/1105.5138}{ArXiv:1105.5138}.

\bibitem[{\citenamefont{Son and Yamamoto}(2012)}]{Son:12:2}
\bibinfo{author}{\bibfnamefont{D.~T.} \bibnamefont{Son}} \bibnamefont{and}
  \bibinfo{author}{\bibfnamefont{N.}~\bibnamefont{Yamamoto}},
  \bibinfo{journal}{Phys.Rev.Lett.} \textbf{\bibinfo{volume}{109}},
  \bibinfo{pages}{181602} (\bibinfo{year}{2012}), \href{http://arxiv.org/abs/1203.2697}{ArXiv:1203.2697}.

\bibitem[{\citenamefont{Stephanov and Yin}(2012)}]{Stephanov:12:1}
\bibinfo{author}{\bibfnamefont{M.~A.} \bibnamefont{Stephanov}}
  \bibnamefont{and} \bibinfo{author}{\bibfnamefont{Y.}~\bibnamefont{Yin}},
  \bibinfo{journal}{Phys.Rev.Lett.} \textbf{\bibinfo{volume}{109}},
  \bibinfo{pages}{162001} (\bibinfo{year}{2012}), \href{http://arxiv.org/abs/1207.0747}{ArXiv:1207.0747}.

\bibitem[{\citenamefont{Son and Yamamoto}(2013)}]{Son:13:1}
\bibinfo{author}{\bibfnamefont{D.~T.} \bibnamefont{Son}} \bibnamefont{and}
  \bibinfo{author}{\bibfnamefont{N.}~\bibnamefont{Yamamoto}},
  \bibinfo{journal}{Phys.Rev.D} \textbf{\bibinfo{volume}{87}},
  \bibinfo{pages}{085016} (\bibinfo{year}{2013}), \href{http://arxiv.org/abs/1210.8158}{ArXiv:1210.8158}.

\bibitem[{\citenamefont{Jensen}(2012)}]{Jensen:12:1}
\bibinfo{author}{\bibfnamefont{K.}~\bibnamefont{Jensen}},
  \bibinfo{journal}{Phys.Rev.D} \textbf{\bibinfo{volume}{85}},
  \bibinfo{pages}{125017} (\bibinfo{year}{2012}), \href{http://arxiv.org/abs/1203.3599}{ArXiv:1203.3599}.

\bibitem[{\citenamefont{Jensen et~al.}(2012)\citenamefont{Jensen, Kaminski,
  Kovtun, Meyer, Ritz, and Yarom}}]{Jensen:12:2}
\bibinfo{author}{\bibfnamefont{K.}~\bibnamefont{Jensen}},
  \bibinfo{author}{\bibfnamefont{M.}~\bibnamefont{Kaminski}},
  \bibinfo{author}{\bibfnamefont{P.}~\bibnamefont{Kovtun}},
  \bibinfo{author}{\bibfnamefont{R.}~\bibnamefont{Meyer}},
  \bibinfo{author}{\bibfnamefont{A.}~\bibnamefont{Ritz}}, \bibnamefont{and}
  \bibinfo{author}{\bibfnamefont{A.}~\bibnamefont{Yarom}},
  \bibinfo{journal}{Phys.Rev.Lett.} \textbf{\bibinfo{volume}{109}},
  \bibinfo{pages}{101601} (\bibinfo{year}{2012}), \href{http://arxiv.org/abs/1203.3556}{ArXiv:1203.3556}.

\bibitem[{\citenamefont{Banerjee et~al.}(2012)\citenamefont{Banerjee,
  Bhattacharya, Bhattacharyya, Jain, Minwalla, and Sharma}}]{Banerjee:12:1}
\bibinfo{author}{\bibfnamefont{N.}~\bibnamefont{Banerjee}},
  \bibinfo{author}{\bibfnamefont{J.}~\bibnamefont{Bhattacharya}},
  \bibinfo{author}{\bibfnamefont{S.}~\bibnamefont{Bhattacharyya}},
  \bibinfo{author}{\bibfnamefont{S.}~\bibnamefont{Jain}},
  \bibinfo{author}{\bibfnamefont{S.}~\bibnamefont{Minwalla}}, \bibnamefont{and}
  \bibinfo{author}{\bibfnamefont{T.}~\bibnamefont{Sharma}},
  \bibinfo{journal}{JHEP} \textbf{\bibinfo{volume}{09}}, \bibinfo{pages}{46}
  (\bibinfo{year}{2012}), \href{http://arxiv.org/abs/1203.3544}{ArXiv:1203.3544}.

\bibitem[{\citenamefont{Buividovich}(2014)}]{Buividovich:13:8}
\bibinfo{author}{\bibfnamefont{P.~V.} \bibnamefont{Buividovich}},
  \bibinfo{journal}{Nucl. Phys. A} \textbf{\bibinfo{volume}{925}},
  \bibinfo{pages}{218 } (\bibinfo{year}{2014}), \href{http://arxiv.org/abs/1312.1843}{ArXiv:1312.1843}.

\bibitem[{\citenamefont{Manuel and
  Torres-Rincon}(2014{\natexlab{a}})}]{TorresRincon:13:1}
\bibinfo{author}{\bibfnamefont{C.}~\bibnamefont{Manuel}} \bibnamefont{and}
  \bibinfo{author}{\bibfnamefont{J.~M.} \bibnamefont{Torres-Rincon}},
  \bibinfo{journal}{Phys.Rev.D} \textbf{\bibinfo{volume}{89}},
  \bibinfo{pages}{096002} (\bibinfo{year}{2014}{\natexlab{a}}),
  \href{http://arxiv.org/abs/1312.1158}{ArXiv:1312.1158}.

\bibitem[{\citenamefont{Manuel and
  Torres-Rincon}(2014{\natexlab{b}})}]{TorresRincon:14:1}
\bibinfo{author}{\bibfnamefont{C.}~\bibnamefont{Manuel}} \bibnamefont{and}
  \bibinfo{author}{\bibfnamefont{J.~M.} \bibnamefont{Torres-Rincon}},
  \bibinfo{journal}{Phys.Rev.D} \textbf{\bibinfo{volume}{90}},
  \bibinfo{pages}{076007} (\bibinfo{year}{2014}{\natexlab{b}}),
  \href{http://arxiv.org/abs/1404.6409}{ArXiv:1404.6409}.

\bibitem[{\citenamefont{Vainshtein}(2003)}]{Vainshtein:03:1}
\bibinfo{author}{\bibfnamefont{A.}~\bibnamefont{Vainshtein}},
  \bibinfo{journal}{Phys.Lett.B} \textbf{\bibinfo{volume}{569}},
  \bibinfo{pages}{187 } (\bibinfo{year}{2003}), \href{http://arxiv.org/abs/hep-ph/0212231}{ArXiv:hep-ph/0212231}.

\bibitem[{\citenamefont{Knecht et~al.}(2004)\citenamefont{Knecht, Peris,
  Perrottet, and {de Rafael}}}]{Knecht:04:1}
\bibinfo{author}{\bibfnamefont{M.}~\bibnamefont{Knecht}},
  \bibinfo{author}{\bibfnamefont{S.}~\bibnamefont{Peris}},
  \bibinfo{author}{\bibfnamefont{M.}~\bibnamefont{Perrottet}},
  \bibnamefont{and} \bibinfo{author}{\bibfnamefont{E.}~\bibnamefont{{de
  Rafael}}}, \bibinfo{journal}{JHEP} \textbf{\bibinfo{volume}{0403}},
  \bibinfo{pages}{035} (\bibinfo{year}{2004}), \href{http://arxiv.org/abs/hep-ph/0311100}{ArXiv:hep-ph/0311100}.

\bibitem[{\citenamefont{Fukushima and Mameda}(2012)}]{Fukushima:12:1}
\bibinfo{author}{\bibfnamefont{K.}~\bibnamefont{Fukushima}} \bibnamefont{and}
  \bibinfo{author}{\bibfnamefont{K.}~\bibnamefont{Mameda}},
  \bibinfo{journal}{Phys.Rev.D} \textbf{\bibinfo{volume}{86}},
  \bibinfo{pages}{071501} (\bibinfo{year}{2012}), \href{http://arxiv.org/abs/1206.3128}{ArXiv:1206.3128}.

\bibitem[{\citenamefont{Kalaydzhyan}(2014)}]{Kalaydzhyan:14:1}
\bibinfo{author}{\bibfnamefont{T.}~\bibnamefont{Kalaydzhyan}},
  \bibinfo{journal}{Phys.Rev.D} \textbf{\bibinfo{volume}{89}},
  \bibinfo{pages}{105012} (\bibinfo{year}{2014}), \href{http://arxiv.org/abs/1403.1256}{ArXiv:1403.1256}.

\bibitem[{\citenamefont{Nair et~al.}(2012)\citenamefont{Nair, Ray, and
  Roy}}]{Ray:12:1}
\bibinfo{author}{\bibfnamefont{V.~P.} \bibnamefont{Nair}},
  \bibinfo{author}{\bibfnamefont{R.}~\bibnamefont{Ray}}, \bibnamefont{and}
  \bibinfo{author}{\bibfnamefont{S.}~\bibnamefont{Roy}},
  \bibinfo{journal}{Phys.Rev.D} \textbf{\bibinfo{volume}{86}},
  \bibinfo{pages}{025012} (\bibinfo{year}{2012}), \href{http://arxiv.org/abs/1112.4022}{ArXiv:1112.4022}.

\bibitem[{\citenamefont{Jensen et~al.}(2013)\citenamefont{Jensen, Kovtun, and
  Ritz}}]{Jensen:13:1}
\bibinfo{author}{\bibfnamefont{K.}~\bibnamefont{Jensen}},
  \bibinfo{author}{\bibfnamefont{P.}~\bibnamefont{Kovtun}}, \bibnamefont{and}
  \bibinfo{author}{\bibfnamefont{A.}~\bibnamefont{Ritz}},
  \bibinfo{journal}{JHEP} \textbf{\bibinfo{volume}{1310}}, \bibinfo{pages}{186}
  (\bibinfo{year}{2013}), \href{http://arxiv.org/abs/1307.3234}{ArXiv:1307.3234}.

\bibitem[{\citenamefont{Gorbar et~al.}(2013)\citenamefont{Gorbar, Miransky,
  Shovkovy, and Wang}}]{Miransky:13:1}
\bibinfo{author}{\bibfnamefont{E.~V.} \bibnamefont{Gorbar}},
  \bibinfo{author}{\bibfnamefont{V.~A.} \bibnamefont{Miransky}},
  \bibinfo{author}{\bibfnamefont{I.~A.} \bibnamefont{Shovkovy}},
  \bibnamefont{and} \bibinfo{author}{\bibfnamefont{X.}~\bibnamefont{Wang}},
  \bibinfo{journal}{Phys.Rev.D} \textbf{\bibinfo{volume}{88}},
  \bibinfo{pages}{025025} (\bibinfo{year}{2013}), \href{http://arxiv.org/abs/1304.4606}{ArXiv:1304.4606}.

\bibitem[{\citenamefont{Gursoy and Jansen}(2014)}]{Gursoy:14:1}
\bibinfo{author}{\bibfnamefont{U.}~\bibnamefont{Gursoy}} \bibnamefont{and}
  \bibinfo{author}{\bibfnamefont{A.}~\bibnamefont{Jansen}},
  \bibinfo{journal}{JHEP} \textbf{\bibinfo{volume}{1410}}, \bibinfo{pages}{92}
  (\bibinfo{year}{2014}), \href{http://arxiv.org/abs/1407.3282}{ArXiv:1407.3282}.

\bibitem[{\citenamefont{Wang and Zhang}(2013)}]{Wang:13:1}
\bibinfo{author}{\bibfnamefont{Z.}~\bibnamefont{Wang}} \bibnamefont{and}
  \bibinfo{author}{\bibfnamefont{S.~C.} \bibnamefont{Zhang}},
  \bibinfo{journal}{Phys.Rev.B} \textbf{\bibinfo{volume}{87}},
  \bibinfo{pages}{161107} (\bibinfo{year}{2013}), \href{http://arxiv.org/abs/1207.5234}{ArXiv:1207.5234}.

\bibitem[{\citenamefont{Wei et~al.}(2012)\citenamefont{Wei, Chao, and
  Aji}}]{Wei:12:1}
\bibinfo{author}{\bibfnamefont{H.}~\bibnamefont{Wei}},
  \bibinfo{author}{\bibfnamefont{S.~P.} \bibnamefont{Chao}}, \bibnamefont{and}
  \bibinfo{author}{\bibfnamefont{V.}~\bibnamefont{Aji}},
  \bibinfo{journal}{Phys.Rev.Lett.} \textbf{\bibinfo{volume}{109}},
  \bibinfo{pages}{196403} (\bibinfo{year}{2012}), \href{http://arxiv.org/abs/1207.5065}{ArXiv:1207.5065}.

\bibitem[{\citenamefont{Sekine and Nomura}(2013)}]{Sekine:13:1}
\bibinfo{author}{\bibfnamefont{A.}~\bibnamefont{Sekine}} \bibnamefont{and}
  \bibinfo{author}{\bibfnamefont{K.}~\bibnamefont{Nomura}},
  \bibinfo{journal}{J.Phys.Soc.Jpn.} \textbf{\bibinfo{volume}{83}},
  \bibinfo{pages}{094710} (\bibinfo{year}{2013}), \href{http://arxiv.org/abs/1309.1079}{ArXiv:1309.1079}.

\bibitem[{\citenamefont{Sukhachov}(2014)}]{Sukhachov:14:1}
\bibinfo{author}{\bibfnamefont{P.~O.} \bibnamefont{Sukhachov}},
  \bibinfo{journal}{Ukr.~J.~Phys.} \textbf{\bibinfo{volume}{59}},
  \bibinfo{pages}{696 } (\bibinfo{year}{2014}), \href{http://arxiv.org/abs/1406.6522}{ArXiv:1406.6522}.

\bibitem[{\citenamefont{Fukushima and Ruggieri}(2010)}]{Fukushima:10:1}
\bibinfo{author}{\bibfnamefont{K.}~\bibnamefont{Fukushima}} \bibnamefont{and}
  \bibinfo{author}{\bibfnamefont{M.}~\bibnamefont{Ruggieri}},
  \bibinfo{journal}{Phys.Rev.D} \textbf{\bibinfo{volume}{82}},
  \bibinfo{pages}{054001} (\bibinfo{year}{2010}), \href{http://arxiv.org/abs/1004.2769}{ArXiv:1004.2769}.

\bibitem[{\citenamefont{Fukushima et~al.}(2010)\citenamefont{Fukushima,
  Ruggieri, and Gatto}}]{Fukushima:10:2}
\bibinfo{author}{\bibfnamefont{K.}~\bibnamefont{Fukushima}},
  \bibinfo{author}{\bibfnamefont{M.}~\bibnamefont{Ruggieri}}, \bibnamefont{and}
  \bibinfo{author}{\bibfnamefont{R.}~\bibnamefont{Gatto}},
  \bibinfo{journal}{Phys.Rev.D} \textbf{\bibinfo{volume}{81}},
  \bibinfo{pages}{114031} (\bibinfo{year}{2010}), \href{http://arxiv.org/abs/1003.0047}{ArXiv:1003.0047}.

\bibitem[{\citenamefont{Gatto and Ruggieri}(2012)}]{Gatto:11:1}
\bibinfo{author}{\bibfnamefont{R.}~\bibnamefont{Gatto}} \bibnamefont{and}
  \bibinfo{author}{\bibfnamefont{M.}~\bibnamefont{Ruggieri}},
  \bibinfo{journal}{Phys.Rev.D} \textbf{\bibinfo{volume}{85}},
  \bibinfo{pages}{054013} (\bibinfo{year}{2012}), \href{http://arxiv.org/abs/1110.4904}{ArXiv:1110.4904}.

\bibitem[{\citenamefont{Ruggieri}(2011)}]{Ruggieri:11:1}
\bibinfo{author}{\bibfnamefont{M.}~\bibnamefont{Ruggieri}},
  \bibinfo{journal}{Phys.Rev.D} \textbf{\bibinfo{volume}{84}},
  \bibinfo{pages}{014011} (\bibinfo{year}{2011}), \href{http://arxiv.org/abs/1103.6186}{ArXiv:1103.6186}.

\bibitem[{\citenamefont{Chernodub and Nedelin}(2011)}]{Nedelin:11:1}
\bibinfo{author}{\bibfnamefont{M.~N.} \bibnamefont{Chernodub}}
  \bibnamefont{and} \bibinfo{author}{\bibfnamefont{A.~S.}
  \bibnamefont{Nedelin}}, \bibinfo{journal}{Phys.Rev.D}
  \textbf{\bibinfo{volume}{83}}, \bibinfo{pages}{105008}
  (\bibinfo{year}{2011}), \href{http://arxiv.org/abs/1102.0188}{ArXiv:1102.0188}.

\bibitem[{\citenamefont{Andrianov
  et~al.}(2014{\natexlab{a}})\citenamefont{Andrianov, Espriu, and
  Planells}}]{Andrianov:13:1}
\bibinfo{author}{\bibfnamefont{A.~A.} \bibnamefont{Andrianov}},
  \bibinfo{author}{\bibfnamefont{D.}~\bibnamefont{Espriu}}, \bibnamefont{and}
  \bibinfo{author}{\bibfnamefont{X.}~\bibnamefont{Planells}},
  \bibinfo{journal}{Eur.Phys.J.C} \textbf{\bibinfo{volume}{74}},
  \bibinfo{pages}{2776} (\bibinfo{year}{2014}{\natexlab{a}}),
  \href{http://arxiv.org/abs/1310.4416}{ArXiv:1310.4416}.

\bibitem[{\citenamefont{Andrianov
  et~al.}(2014{\natexlab{b}})\citenamefont{Andrianov, Andrianov, Espriu, and
  Planells}}]{Andrianov:13:2}
\bibinfo{author}{\bibfnamefont{A.~A.} \bibnamefont{Andrianov}},
  \bibinfo{author}{\bibfnamefont{V.~A.} \bibnamefont{Andrianov}},
  \bibinfo{author}{\bibfnamefont{D.}~\bibnamefont{Espriu}}, \bibnamefont{and}
  \bibinfo{author}{\bibfnamefont{X.}~\bibnamefont{Planells}},
  \bibinfo{journal}{PoS} \textbf{\bibinfo{volume}{QFTHEP2013}},
  \bibinfo{pages}{049} (\bibinfo{year}{2014}{\natexlab{b}}),
  \href{http://arxiv.org/abs/1310.4434}{ArXiv:1310.4434}.

\bibitem[{\citenamefont{{Ballon Bayona} et~al.}(2011)\citenamefont{{Ballon
  Bayona}, Peeters, and Zamaklar}}]{Zamaklar:11:1}
\bibinfo{author}{\bibfnamefont{C.~A.} \bibnamefont{{Ballon Bayona}}},
  \bibinfo{author}{\bibfnamefont{K.}~\bibnamefont{Peeters}}, \bibnamefont{and}
  \bibinfo{author}{\bibfnamefont{M.}~\bibnamefont{Zamaklar}},
  \bibinfo{journal}{JHEP} \textbf{\bibinfo{volume}{1106}}, \bibinfo{pages}{092}
  (\bibinfo{year}{2011}), \href{http://arxiv.org/abs/1104.2291}{ArXiv:1104.2291}.

\bibitem[{\citenamefont{Akamatsu and Yamamoto}(2013)}]{Yamamoto:13:1}
\bibinfo{author}{\bibfnamefont{Y.}~\bibnamefont{Akamatsu}} \bibnamefont{and}
  \bibinfo{author}{\bibfnamefont{N.}~\bibnamefont{Yamamoto}},
  \bibinfo{journal}{Phys.Rev.Lett.} \textbf{\bibinfo{volume}{111}},
  \bibinfo{pages}{052002} (\bibinfo{year}{2013}), \href{http://arxiv.org/abs/1302.2125}{ArXiv:1302.2125}.

\bibitem[{\citenamefont{Khaidukov et~al.}(2013)\citenamefont{Khaidukov,
  Kirilin, Sadofyev, and Zakharov}}]{Sadofyev:13:1}
\bibinfo{author}{\bibfnamefont{Z.~V.} \bibnamefont{Khaidukov}},
  \bibinfo{author}{\bibfnamefont{V.~P.} \bibnamefont{Kirilin}},
  \bibinfo{author}{\bibfnamefont{A.~V.} \bibnamefont{Sadofyev}},
  \bibnamefont{and} \bibinfo{author}{\bibfnamefont{V.~I.}
  \bibnamefont{Zakharov}}, \emph{On magnetostatics of chiral
  media} (\bibinfo{year}{2013}), \href{http://arxiv.org/abs/1307.0138}{ArXiv:1307.0138}.

\bibitem[{\citenamefont{Parameswaran et~al.}(2014)\citenamefont{Parameswaran,
  Grover, Abanin, Pesin, and Vishwanath}}]{Parameswaran:13:1}
\bibinfo{author}{\bibfnamefont{S.~A.} \bibnamefont{Parameswaran}},
  \bibinfo{author}{\bibfnamefont{T.}~\bibnamefont{Grover}},
  \bibinfo{author}{\bibfnamefont{D.~A.} \bibnamefont{Abanin}},
  \bibinfo{author}{\bibfnamefont{D.~A.} \bibnamefont{Pesin}}, \bibnamefont{and}
  \bibinfo{author}{\bibfnamefont{A.}~\bibnamefont{Vishwanath}},
  \bibinfo{journal}{Phys.Rev.X} \textbf{\bibinfo{volume}{4}},
  \bibinfo{pages}{031035} (\bibinfo{year}{2014}), \href{http://arxiv.org/abs/1306.1234}{ArXiv:1306.1234}.

\bibitem[{\citenamefont{Ashby and Carbotte}(2014)}]{Ashby:14:1}
\bibinfo{author}{\bibfnamefont{P.~E.~C.} \bibnamefont{Ashby}} \bibnamefont{and}
  \bibinfo{author}{\bibfnamefont{J.~P.} \bibnamefont{Carbotte}},
  \bibinfo{journal}{Phys.Rev.B} \textbf{\bibinfo{volume}{89}},
  \bibinfo{pages}{245121} (\bibinfo{year}{2014}), \href{http://arxiv.org/abs/1405.7034}{ArXiv:1405.7034}.

\bibitem[{\citenamefont{Hosur and Qi}(2014)}]{Hosur:14:1}
\bibinfo{author}{\bibfnamefont{P.}~\bibnamefont{Hosur}} \bibnamefont{and}
  \bibinfo{author}{\bibfnamefont{X.}~\bibnamefont{Qi}},
  \emph{Tunable optical activity due to the chiral anomaly in
  {Weyl} semimetals} (\bibinfo{year}{2014}), \href{http://arxiv.org/abs/1401.2762}{ArXiv:1401.2762}.

\bibitem[{\citenamefont{Altland and Simons}(2010)}]{AltlandSimonsCondMatQFT}
\bibinfo{author}{\bibfnamefont{A.}~\bibnamefont{Altland}} \bibnamefont{and}
  \bibinfo{author}{\bibfnamefont{B.~D.} \bibnamefont{Simons}},
  \emph{Condensed Matter Field Theory}
  (\bibinfo{publisher}{Cambridge University Press}, \bibinfo{year}{2010}).

\bibitem[{\citenamefont{Buividovich and Puhr}(2014)}]{Buividovich:14:2}
\bibinfo{author}{\bibfnamefont{P.~V.} \bibnamefont{Buividovich}}
  \bibnamefont{and} \bibinfo{author}{\bibfnamefont{M.}~\bibnamefont{Puhr}},
  \bibinfo{journal}{PoS} \textbf{\bibinfo{volume}{Lattice2014}},
  \bibinfo{pages}{061} (\bibinfo{year}{2014}), \href{http://arxiv.org/abs/1410.6704}{ArXiv:1410.6704}.

\bibitem[{\citenamefont{Gynther et~al.}(2011)\citenamefont{Gynther,
  Landsteiner, {Pena-Benitez}, and Rebhan}}]{Gynther:10:1}
\bibinfo{author}{\bibfnamefont{A.}~\bibnamefont{Gynther}},
  \bibinfo{author}{\bibfnamefont{K.}~\bibnamefont{Landsteiner}},
  \bibinfo{author}{\bibfnamefont{F.}~\bibnamefont{{Pena-Benitez}}},
  \bibnamefont{and} \bibinfo{author}{\bibfnamefont{A.}~\bibnamefont{Rebhan}},
  \bibinfo{journal}{JHEP} \textbf{\bibinfo{volume}{1102}}, \bibinfo{pages}{110}
  (\bibinfo{year}{2011}), \href{http://arxiv.org/abs/1005.2587}{ArXiv:1005.2587}.

\bibitem[{\citenamefont{Amado et~al.}(2011)\citenamefont{Amado, Landsteiner,
  and Pena-Benitez}}]{Landsteiner:11:2}
\bibinfo{author}{\bibfnamefont{I.}~\bibnamefont{Amado}},
  \bibinfo{author}{\bibfnamefont{K.}~\bibnamefont{Landsteiner}},
  \bibnamefont{and}
  \bibinfo{author}{\bibfnamefont{F.}~\bibnamefont{Pena-Benitez}},
  \bibinfo{journal}{JHEP} \textbf{\bibinfo{volume}{05}}, \bibinfo{pages}{081}
  (\bibinfo{year}{2011}), \href{http://arxiv.org/abs/1102.4577}{ArXiv:1102.4577}.

\bibitem[{\citenamefont{Landsteiner et~al.}(2012)\citenamefont{Landsteiner,
  Megias, and {Pena-Benitez}}}]{Landsteiner:12:1}
\bibinfo{author}{\bibfnamefont{K.}~\bibnamefont{Landsteiner}},
  \bibinfo{author}{\bibfnamefont{E.}~\bibnamefont{Megias}}, \bibnamefont{and}
  \bibinfo{author}{\bibfnamefont{F.}~\bibnamefont{{Pena-Benitez}}},
  \emph{Anomalous transport from {Kubo} formulae},
  \bibinfo{howpublished}{in Lect. Notes Phys. {Strongly interacting matter in
  magnetic fields} (Springer), edited by D. Kharzeev, K. Landsteiner, A.
  Schmitt, H.-U. Yee} (\bibinfo{year}{2012}), \href{http://arxiv.org/abs/1207.5808}{ArXiv:1207.5808}.

\bibitem[{\citenamefont{Rebhan et~al.}(2010)\citenamefont{Rebhan, Schmitt, and
  Stricker}}]{Rebhan:10:1}
\bibinfo{author}{\bibfnamefont{A.}~\bibnamefont{Rebhan}},
  \bibinfo{author}{\bibfnamefont{A.}~\bibnamefont{Schmitt}}, \bibnamefont{and}
  \bibinfo{author}{\bibfnamefont{S.~A.} \bibnamefont{Stricker}},
  \bibinfo{journal}{JHEP} \textbf{\bibinfo{volume}{1001}}, \bibinfo{pages}{026}
  (\bibinfo{year}{2010}), \href{http://arxiv.org/abs/0909.4782}{ArXiv:0909.4782}.

\bibitem[{\citenamefont{Rubakov}(2010)}]{Rubakov:10:1}
\bibinfo{author}{\bibfnamefont{V.~A.} \bibnamefont{Rubakov}},
  \emph{On chiral magnetic effect and holography}
  (\bibinfo{year}{2010}), \href{http://arxiv.org/abs/1005.1888}{ArXiv:1005.1888}.

\bibitem[{\citenamefont{Kharzeev and Warringa}(2009)}]{Kharzeev:09:1}
\bibinfo{author}{\bibfnamefont{D.~E.} \bibnamefont{Kharzeev}} \bibnamefont{and}
  \bibinfo{author}{\bibfnamefont{H.~J.} \bibnamefont{Warringa}},
  \bibinfo{journal}{Phys.Rev.D} \textbf{\bibinfo{volume}{80}},
  \bibinfo{pages}{034028} (\bibinfo{year}{2009}), \href{http://arxiv.org/abs/0907.5007}{ArXiv:0907.5007}.

\bibitem[{\citenamefont{Zyuzin and Burkov}(2012)}]{Zyuzin:12:2}
\bibinfo{author}{\bibfnamefont{A.~A.} \bibnamefont{Zyuzin}} \bibnamefont{and}
  \bibinfo{author}{\bibfnamefont{A.~A.} \bibnamefont{Burkov}},
  \bibinfo{journal}{Phys.Rev.B} \textbf{\bibinfo{volume}{86}},
  \bibinfo{pages}{115133} (\bibinfo{year}{2012}), \href{http://arxiv.org/abs/1206.1868}{ArXiv:1206.1868}.

\bibitem[{\citenamefont{Buividovich}(2013)}]{Buividovich:13:6}
\bibinfo{author}{\bibfnamefont{P.~V.} \bibnamefont{Buividovich}},
  \bibinfo{journal}{PoS} \textbf{\bibinfo{volume}{LATTICE2013}},
  \bibinfo{pages}{179} (\bibinfo{year}{2013}), \href{http://arxiv.org/abs/1309.2850}{ArXiv:1309.2850}.

\bibitem[{\citenamefont{Chao et~al.}(2013)\citenamefont{Chao, Chu, and
  Huang}}]{Chao:13:1}
\bibinfo{author}{\bibfnamefont{J.}~\bibnamefont{Chao}},
  \bibinfo{author}{\bibfnamefont{P.}~\bibnamefont{Chu}}, \bibnamefont{and}
  \bibinfo{author}{\bibfnamefont{M.}~\bibnamefont{Huang}},
  \bibinfo{journal}{Phys.Rev.D} \textbf{\bibinfo{volume}{88}},
  \bibinfo{pages}{054009} (\bibinfo{year}{2013}), \href{http://arxiv.org/abs/1305.1100}{ArXiv:1305.1100}.

\bibitem[{\citenamefont{Yu et~al.}(2014)\citenamefont{Yu, Liu, and
  Huang}}]{Yu:14:1}
\bibinfo{author}{\bibfnamefont{L.}~\bibnamefont{Yu}},
  \bibinfo{author}{\bibfnamefont{H.}~\bibnamefont{Liu}}, \bibnamefont{and}
  \bibinfo{author}{\bibfnamefont{M.}~\bibnamefont{Huang}},
  \bibinfo{journal}{Phys.Rev.D} \textbf{\bibinfo{volume}{90}},
  \bibinfo{pages}{074009} (\bibinfo{year}{2014}), \href{http://arxiv.org/abs/1404.6969}{ArXiv:1404.6969}.

\bibitem[{\citenamefont{Hosur and Qi}(2013)}]{Hosur:13:1}
\bibinfo{author}{\bibfnamefont{P.}~\bibnamefont{Hosur}} \bibnamefont{and}
  \bibinfo{author}{\bibfnamefont{X.}~\bibnamefont{Qi}},
  \bibinfo{journal}{Comp.Rend.Phys.} \textbf{\bibinfo{volume}{14}},
  \bibinfo{pages}{857 } (\bibinfo{year}{2013}), \href{http://arxiv.org/abs/1309.4464}{ArXiv:1309.4464}.

\bibitem[{\citenamefont{Creutz et~al.}(2002)\citenamefont{Creutz, {Horvath},
  and Neuberger}}]{Creutz:01:1}
\bibinfo{author}{\bibfnamefont{M.}~\bibnamefont{Creutz}},
  \bibinfo{author}{\bibfnamefont{I.}~\bibnamefont{{Horvath}}},
  \bibnamefont{and}
  \bibinfo{author}{\bibfnamefont{H.}~\bibnamefont{Neuberger}},
  \bibinfo{journal}{Nucl.Phys.Proc.Suppl.} \textbf{\bibinfo{volume}{106}},
  \bibinfo{pages}{760} (\bibinfo{year}{2002}), \href{http://arxiv.org/abs/hep-lat/0110009}{ArXiv:hep-lat/0110009}.

\bibitem[{\citenamefont{Aoki}(1984)}]{Aoki:84:1}
\bibinfo{author}{\bibfnamefont{S.}~\bibnamefont{Aoki}},
  \href{http://link.aps.org/doi/10.1103/PhysRevD.30.2653}{
  \bibinfo{journal}{Phys.Rev.D} \textbf{\bibinfo{volume}{30}},
  \bibinfo{pages}{2653 } (\bibinfo{year}{1984})}.

\bibitem[{\citenamefont{Yamamoto}(2011)}]{Yamamoto:11:1}
\bibinfo{author}{\bibfnamefont{A.}~\bibnamefont{Yamamoto}},
  \bibinfo{journal}{Phys.Rev.Lett.} \textbf{\bibinfo{volume}{107}},
  \bibinfo{pages}{031601} (\bibinfo{year}{2011}), \href{http://arxiv.org/abs/1105.0385}{ArXiv:1105.0385}.

\bibitem[{\citenamefont{Montvay and Muenster}(1994)}]{MontvayMuenster}
\bibinfo{author}{\bibfnamefont{I.}~\bibnamefont{Montvay}} \bibnamefont{and}
  \bibinfo{author}{\bibfnamefont{G.}~\bibnamefont{Muenster}},
  \emph{Quantum fields on a lattice}
  (\bibinfo{publisher}{Cambridge University Press}, \bibinfo{year}{1994}).

\bibitem[{\citenamefont{Frolov and Slavnov}(1993)}]{Slavnov:93:1}
\bibinfo{author}{\bibfnamefont{S.~A.} \bibnamefont{Frolov}} \bibnamefont{and}
  \bibinfo{author}{\bibfnamefont{A.~A.} \bibnamefont{Slavnov}},
  \href{http://dx.doi.org/10.1016/0370-2693(93)90943-C}{\bibinfo{journal}{Phys.Lett.B} \textbf{\bibinfo{volume}{309}}, \bibinfo{pages}{344 } (\bibinfo{year}{1993})}.

\bibitem[{\citenamefont{Narayanan and Neuberger}(1993)}]{Neuberger:93:1}
\bibinfo{author}{\bibfnamefont{R.}~\bibnamefont{Narayanan}} \bibnamefont{and}
  \bibinfo{author}{\bibfnamefont{H.}~\bibnamefont{Neuberger}},
  \bibinfo{journal}{Phys.Lett.B} \textbf{\bibinfo{volume}{302}},
  \bibinfo{pages}{62 } (\bibinfo{year}{1993}), \href{http://arxiv.org/abs/hep-lat/9212019}{ArXiv:hep-lat/9212019}.

\bibitem[{\citenamefont{Johnson}()}]{Johnson:Cubature}
\bibinfo{author}{\bibfnamefont{S.~G.} \bibnamefont{Johnson}},
  \emph{Cubature (multi-dimensional integration)},
  \href{http://ab-initio.mit.edu/wiki/index.php/Cubature}{http://ab-initio.mit.edu/wiki/index.php/Cubature}.

\end{thebibliography}

\end{document}